	\documentclass[preprint]{aastex}

	\newcommand{\modification}[1]{#1}

	\newcommand{\gband}{{\rm G-band}}
	\newcommand{\misma}{{MISMA}}

	\newcommand{\citeN}[1]{\citeauthor{#1} (\citeyear{#1})}
	\newcommand{\citeNP}[1]{\citeauthor{#1} \citeyear{#1}}

	\shorttitle{Sinthesis of solar \gband\ spectra}
\begin{document}

\title{G-band Spectral Synthesis in Solar Magnetic 
	Concentrations}


\author{J. S\'anchez Almeida,
	A. Asensio Ramos,
	and J. Trujillo Bueno\altaffilmark{1}}
\affil{Instituto de Astrof\'\i sica de Canarias, E-38200 La Laguna, Tenerife, 
Spain}
\email{{jos@ll.iac.es}, {aasensio@ll.iac.es}, {jtb@ll.iac.es}}

\and

\author{J. Cernicharo\altaffilmark{1}}
\affil{Instituto de Estructura de la Materia, Serrano 123, E-28006 Madrid, 
Spain}
\email{cerni@astro.iem.csic.es}%cernicharo@argen.net}


\altaffiltext{1}{{Consejo Superior de Investigaciones Cient\'\i ficas, Spain.}}

\begin{abstract}
	Narrow band imaging in the \gband\ is commonly used 
	to trace the small magnetic field concentrations of the 
	Sun, although the mechanism that 
	makes them bright has remained unclear. We carry
	out LTE syntheses of the \gband\ in an assorted set of
	semi-empirical model magnetic concentrations.
	The syntheses include all CH lines as well as the
	main atomic lines within the band-pass.
	The model atmospheres produce bright  \gband\ spectra having many
	properties in common with the observed \gband\ bright points.
	In particular, the contrast referred to the quiet Sun is
	about twice the contrast in continuum wavelengths.
	The agreement with observations
	does not depend on the specificities of the model atmosphere,
	rather it holds from
	single fluxtubes to MIcro-Structured Magnetic Atmospheres.
	However, the agreement  requires
	that the real \gband\ bright points are not spatially
	resolved,
	even in the best observations.
	Since the predicted \gband\ intensities 
	exceed by far the  observed values, 
	we foresee a notable increase of contrast of the 
	\gband\ images upon
	improvement of the angular resolution.
	According to the LTE modeling,
	the \gband\ spectrum emerges from the deep photosphere that
	produces the
	continuum. Our syntheses also
	predict
	solar magnetic concentrations showing up in continuum images
	but not in the
	\gband . Finally, we
	have examined the importance of the CH photo-dissociation 
	in setting the amount of \gband\ absorption. It turns out
	to play a minor role.
\end{abstract}


%%% From the front matter, we move on to the body of the paper.
\newpage
\section{Introduction\label{introduction}}

	Most of the solar magnetic structures are far too
small to be spatially resolved with the current instruments.
Although this limitation severely hampers any observational
study,
we have all sorts of good reasons to try to
find out their properties. 
Depending on several not yet known attributes
(fraction of the Sun that they cover, 
degree of concentration, grade of tangling, etc),
unresolved elusive magnetic structures
may contain nearly all 
the magnetic flux and energy of the solar photosphere 
	(e.g., \citeNP{ste77};
	\citeNP{ste82};
	\citeNP{yi93};
	\citeNP{san98c};
	\citeNP{san00}; 
	\citeNP{san00c}).
	Should this be the case,
	a proper description of
the solar magnetism relies on a correct
characterization  of these  seemingly second-rate features.

The physical properties of these structures have to be investigated 
using techniques that circumvent the lack of resolution.
Two approaches
have been traditionally explored. First, the measurement and 
careful interpretation of the line
polarization. This observable minimizes the contamination of
the non-magnetic structures existing in the resolution
element, which produce no signal. Since the line polarization is always 
extremely low, the measurements require long integration times that
compromise the spatial resolution.
The second approach uses (broad-band) imaging at selected wavelengths.
In this case the contrast of the non-magnetic components is not 
negligible, however, the increase of
photon flux allows applying
high angular resolution techniques (e.g., \citeNP{lof98}; \citeNP{bon99}).
These techniques
frequently render images that are 
not limited by seeing, providing information
as close as technically possible to the real structures.
Among the second category,
the use of \gband\ imaging to study the dynamics of the concentrations
has become very successful (\citeNP{ber95}; \citeNP{ber98};
	\citeNP{bal98}).
The \gband\  
results from electronic
transitions of the CH radical at some 4300 \AA , 
being so strong in cool stars as to be part of
the stellar classification criteria.
The association of the point-like brightenings 
observed in solar images with the magnetic concentrations
dates back to the days where the first Bright Points (BPs)
were photographed (\citeNP{dun73}; \citeNP{meh74}).
Observing these concentrations in the \gband\
has clear advantages: they show a particularly large
contrast 
(see, \citeNP{mul84}; \citeNP{ber95}; \citeNP{ber98b}),
and the wavelength is not far from the maximum emission of the solar 
spectrum.
Despite an extensive use during the last decade, the reason or reasons  that
make magnetic concentrations so conspicuous 
in the \gband\ remain unclear (see the introductory sections of
\citeNP{ber96}, and \citeNP{ber98b}, as well as the review by \citeNP{rut99}).
Several possibilities have been advanced in the literature. 
\citeN{ber95} point out the {\it collisional excitation
of CH to the A$^2\Delta$ electronic state by
field aligned currents within the fluxtubes}.
\citeN{rut99} assigns to the conventional wisdom  
{\it that filigree grains brighten in the 
\gband\ because the CH lines cause radiation to 
escape somewhat higher up in the atmosphere
where the fluxtubes are already heated}.
Rutten does not favor this conventional wisdom, so 
he proposes yet another hypothesis:
the CH gets dissociated by the strong radiation field within 
the magnetic fluxtubes.  The photo-dissociation of the CH would reduce
the line opacity within the tube producing a contrast
enhancement.
Finally, \citeauthor{san00} (\citeyear{san00}; \S 4.1)
argue that the LTE spectrum of existing model magnetic concentrations
may account for the behavior of the
\gband , as it also 
explains why the magnetic structures are
bright in continuum.
In this case the CH deficit is controlled by the
hot temperatures of the magnetic photosphere,
and the 
thermalizing collisions in
such dense medium.

Having a good knowledge  on the 
effects that produce the brightness is critical to
fully exploit the diagnostic capabilities of the
\gband\ observations.
It would  allow identifying the 
radiative transfer processes responsible for the
\gband~ enhancement and, therefore, linking  the observable quantities
to the physical properties of the atmosphere
that one would like to retrieve.
Our work aims at testing
some of the mechanisms that have been proposed
to see whether they offer a paradigm  to build on.
Starting from the simplest possibility, we develop 
a LTE synthesis code (\S\S \ref{ch} and \ref{lteformation}), which is then 
used to calculate the \gband\ spectrum in different
model magnetic concentrations (usually derived
from polarimetric data).
This LTE synthesis accounts for the 
main observational facts,  provided the
BPs are still not resolved even in the best observations (\S \ref{comparison}).
The goodness of the LTE modeling seems to be a robust result since it does not
depend on details of the model atmospheres.
We also explore deviations from LTE effects. In particular,
\S \ref{photod} shows the CH photo-dissociation to be of secondary
importance 
under the conditions prevailing in the photosphere.
The physical ingredients that produce the LTE
\gband\ contrast enhancement 
are pointed out in \S \ref{why}.
An orderly discussion on all these findings is
presented in \S \ref{conclusion}.
%%%%%%%%%%%%%%%%%%%%%%%%%
%%%%%%%%%%%%%%%%%%%%%%%%
\section{Model CH molecule\label{ch}}

CH is among the most important molecules in cool stars, and one 
of the few molecules with carbon that can be detected in oxygen-rich stars. CH 
absorption
bands extend from the near ultraviolet ($\lambda$ 2800 \AA) to the mid
infrared ($\lambda\ 5.2\ \mu$m), with very strong contributions
in the visible. We study the so-called \gband\ at some 4300 \AA , which stands out as a  
distinctive spectral feature for
stars from spectral type early F to early M (\citeNP{kee42}; \citeNP{sch82}).
\citeN{jor96}
computed bound-bound transitions and line intensities
for the CH molecule, generating a list of about 113000 lines with their 
oscillator
strengths. This table contains several types of transitions; the three 
electronic
systems \( A^{2}\Delta -X^{2}\Pi  \), \( B^{2}\Sigma ^{-}-X^{2}\Pi  \) and
\( C^{2}\Sigma ^{+}-X^{2}\Pi  \),  and the vibration-rotation bands
of the ground electronic state.
The \gband\  is dominated by transitions of the \( A^{2}\Delta -X^{2}\Pi  
\)
system (\citeNP{jor96}). 

The synthesis of the  \gband\ spectrum requires combining 
several different ingredients. First,
one needs 
the abundance of CH through the atmosphere.  Then, 
the strength of each possible transition within the
spectral region of interest has to be evaluated. By considering the contribution
of the whole set of CH lines, one forms the global 
absorption coefficient that finally enters into the
spectral synthesis. 
We solve the various steps of the problem by generating a 
Thermodynamic Equilibrium (TE) chemical model of CH.
The general procedure can be found 
in the literature 
(see e.g. \citeNP{rus34}; 
	\citeNP{tsu64}; 
	\citeNP{tsu73};  
	\citeNP{hea73};
	\citeNP{tej91}),
however, we will
outline it in \S\S \ref{cha} and \ref{mlines}  to
clarify  
the specificities of our modeling.
In addition, some of theses basic equations of the TE modeling are explicitly
used along the paper (e.g., in \S \ref{photod}
to discuss the importance
of photo-dissociation, and in \S \ref{why} 
to pin down the mechanism responsible for the
\gband\ emission).

\subsection{Solar abundance of CH\label{cha}}
The physical conditions in the solar  photosphere favor the formation of
several diatomic molecules.  Both the photospheric mass densities and the
molecular dissociation energies
are high enough so that some of them are fairly stable
even in such a high energy environment. 
CO is
one of the most abundant molecules, and almost all the carbon which
takes part in molecule formation goes to it. Something very similar happens
with H$_2$, which absorbs almost all the hydrogen 
in molecules. This carbon and hydrogen depletion partly
explains why the CH abundance is lower than the abundance of 
CO and H$_2$. However, 
the dimming of
CH is mostly produced by its small
dissociation energy
(\( D_{0}=3.47 \) eV), being easily dissociated by
low energy collisions and near ultraviolet photons.

\begin{figure}
	%\plotone{/net/mero/scratch/aasensio/textos/jorge/figuras/abund.eps}
	%\plotone{/home/jos/texto/papers/paper32/abund.ps}
	%\plotone{fig1.ps}
	% This new figure includes modifications suggested by the referee
	\plotone{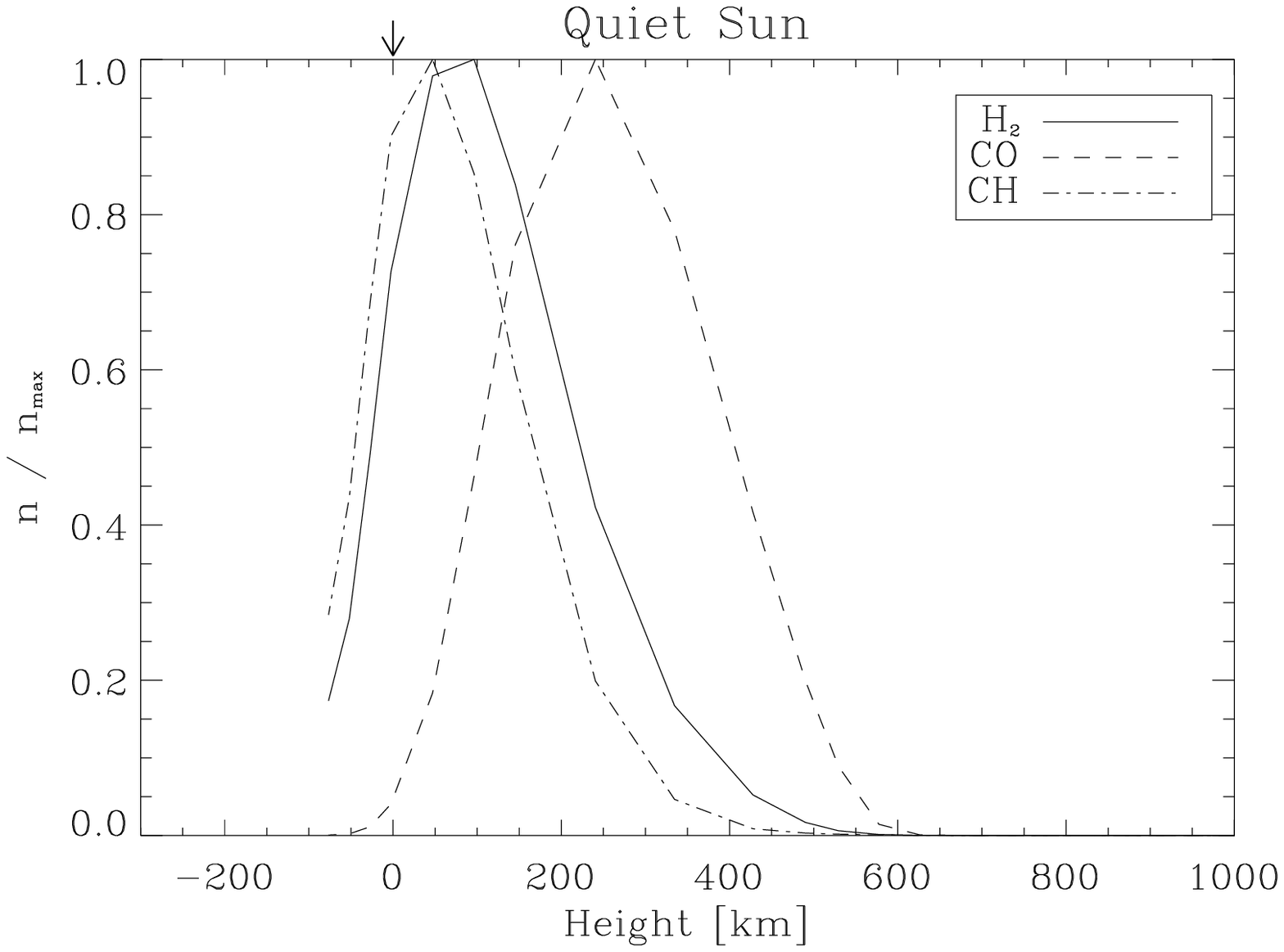}
\caption{Abundance of the main diatomic molecules in the quiet Sun VAL-C model
	atmosphere. The variation with height of each species is 
	normalized to its 
	maximum value: 3.4 $\times\ 10^{13}$  cm$^{-3}$ for H$_2$, 
	2.4 $\times\ 10^{12}$ cm$^{-3}$ for CO, and
	7.2 $\times\ 10^{9}$ cm$^{-3}$ for CH.  Note that the CH peaks 
	deep down in the atmosphere, just some 50 km above 
	continuum optical depth unity at 5000 \AA\ (which sets the zero of the
	abscissa; 
	\modification{
	the small arrow on the top axis points out this height). 
	}
\label{abund}}
\end{figure}

Under TE, the  molecular abundances just depend 
on the temperature and the density. The specific
reaction mechanisms that create and destroy
them are irrelevant.
In spite of the simplification,
TE holds in  many practical situations
(see \S \ref{photod}, where we analyze the case of the
CH chemical evolution in the photosphere).
The number of 
molecules and atoms are coupled 
via the conservation of mass,
and the chemical  equilibrium.
These constraints provide a set of algebraic equations that one solves to
compute the abundances:

\begin{description}
\item - Mass conservation for each constituent, 
\begin{equation}
\label{cons_densidad}
n_{i}+\sum _{m}N_{i}^{m}n_{m}=A_{i}n_{\rm Ht},
\end{equation}
where \( n_{i} \) stands for the atomic abundance (in cm$^{-3}$) of the $i$-th constituent
(atoms in free form, not bounded in molecules), \( A_{i} \) is the total 
abundance
of this element relative to H (we take the solar abundances
by \citeNP{gre84}), \( n_{m} \) represents the molecular abundance (in cm$^{-3}$)
of those molecules having element $i$ in their formulae, \( N_{i}^{m} \) is the
number of nuclei of element $i$ present in molecule $m$ and, finally, \( n_{\rm 
Ht} \) stands for the
number density of H particles (considering all atoms and
molecules in all ionization states; also in cm$^{-3}$).

\item - Saha equation for each molecule (chemical equilibrium for each molecule), 
\begin{equation}
\label{Saha}
\label{cons_densidad1}
n_{m}=\prod _{i}(f^{i}\ n_{i})^{N_{i}^{m}}\phi _{m},
\end{equation}
where \( f^{i} \) is the fraction of $i$ atoms with the
ionization state required to form the molecule $m$,
and \( \phi _{m} \) represents 
the so-called equilibrium constant. The equilibrium constant is actually
a function which, for ideal gases, only depends on temperature $T$. 
\modification{
This approximation is very good for the photospheric
plasma (e.g., \citeNP{cox68}, \S 15.5.)
The
}equilibrium constant has a strong dependence on the dissociation energy,
which we factored for convenience, 
\begin{equation}
\label{Equilibrium_cte}
\phi_m (T)=\phi_m '(T)\exp [D_{0}/(kT)].
\end{equation}
The new factor \( \phi_m '(T) \) can be calculated from molecular
data, including the masses of the elements taking part in the molecular 
formation,
and the partition function of the molecule and its atomic constituents.
The chemical equilibrium constants employed in our calculations
have been taken from \citeN{sau84} and \citeN{tej91}.
As usual, $k$ in equation 
(\ref{Equilibrium_cte}) stands for the Boltzmann constant.
\end{description}
The system of nonlinear equations
(\ref{cons_densidad}) and (\ref{cons_densidad1}) 
can be solved 
to derive the abundances
with any of the usual methods.
We employ Newton-Raphson as described by
\citeN{pre88}.  Since 
the physical conditions change along the atmosphere,
the system 
has to be worked out at each point. 
As an example, the problem for
the VAL-C quiet Sun model atmosphere (\citeNP{ver81}) has been solved and  some
of the results appear in
Figure \ref{abund}. 
The CH abundance has a peak at the bottom of the photosphere, where
the temperature and density conditions allow an efficient formation.
Its abundance is largest at around 50 km, and
it goes away where the CO formation
peaks.

\placefigure{abund}

\begin{figure}
	%\plottwo{/net/mero/scratch/aasensio/textos/jorge/figuras/quiet2.ps}
	%   {/net/mero/scratch/aasensio/textos/jorge/figuras/umbra2.ps}
	%\plottwo{/home/jos/texto/papers/paper32/quiet3.ps}
	%{/home/jos/texto/papers/paper32/umbra3.ps}
%\plottwo{fig2a.ps}{fig2b.ps}
\plottwo{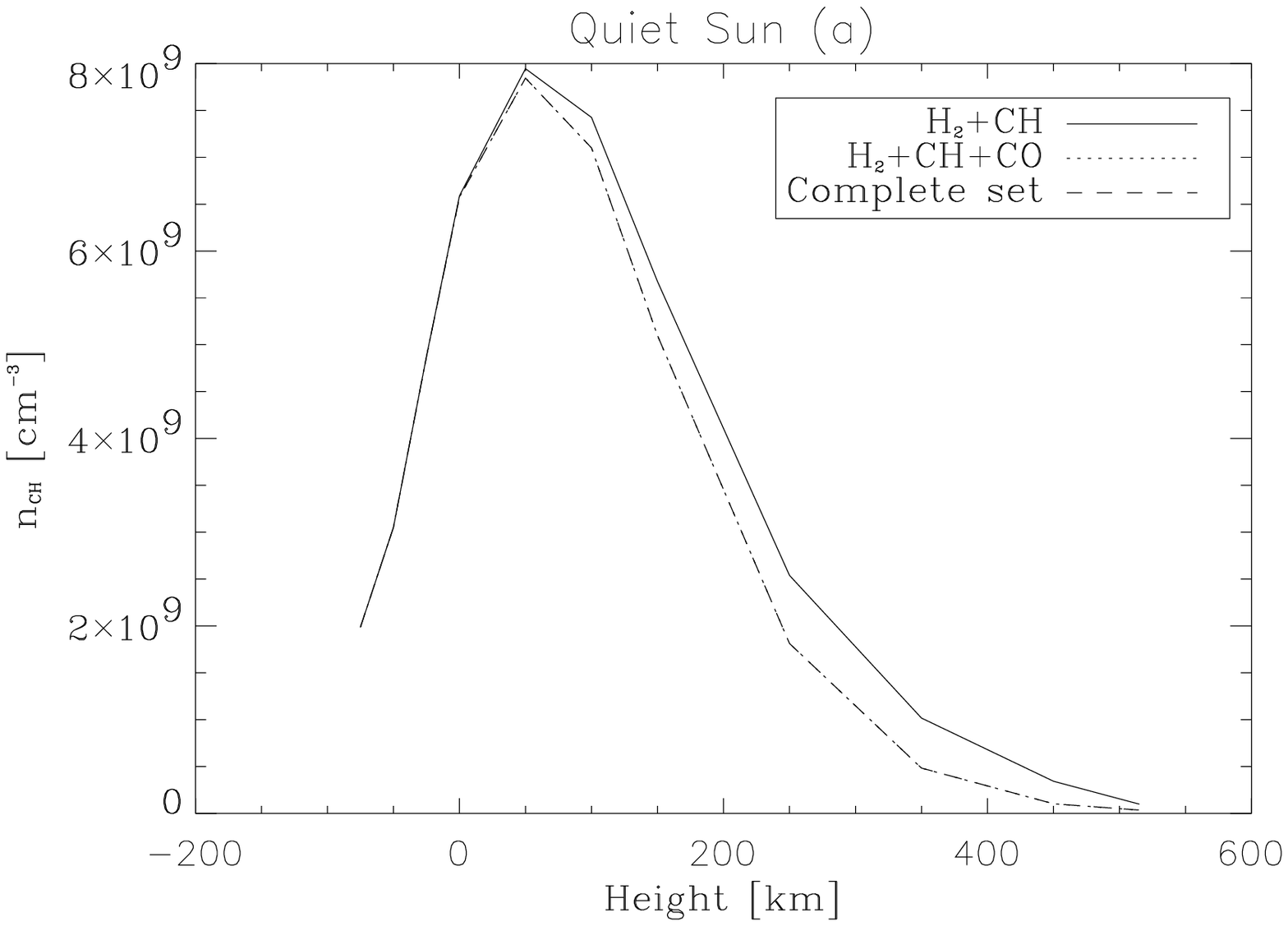}{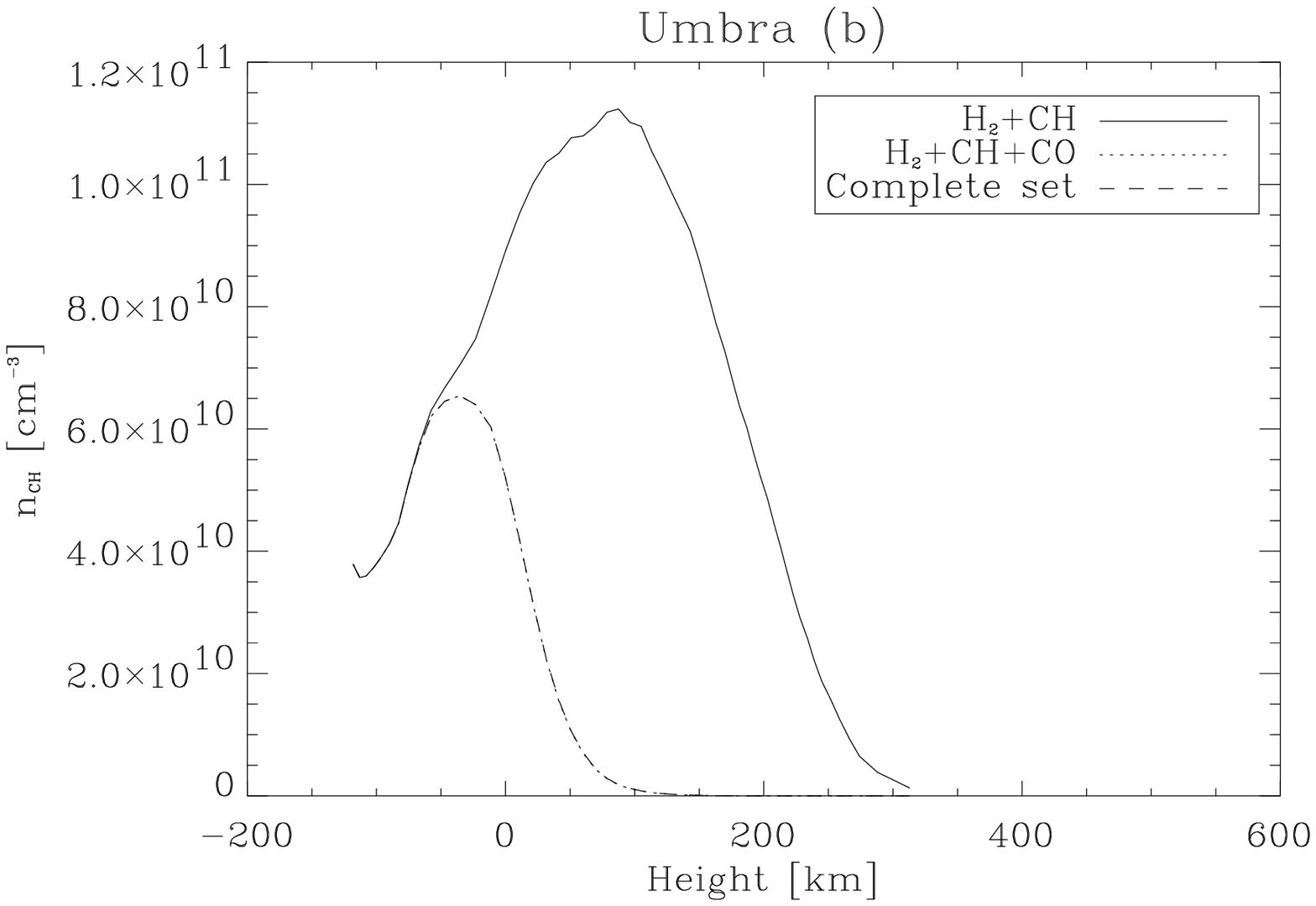}
\caption{CH abundance deduced when considering different sets of 
	molecules  to solve the chemical equilibrium equations:
	H$_2$ and CH (the solid lines), H$_2$, CO and CH (the dotted lines)
	and up to 12 molecular species, including H$_2$, CO, CH (the dashed lines).
	The latter can be regarded as the exact solution.
	In the quiet Sun (a),
	even considering the sole formation of CH gives reasonable
	values. This approximation clearly breaks down in umbrae (b:
	\citeNP{col94} model umbra).
	In this case, however, including just H$_2$ and CO renders good results.
	\label{quiet2}}
\end{figure}

What molecules have to be considered 
to compute the CH abundance?
Except in the umbrae of sunspots, 
most of the photospheric plasma is in the form of atoms, consequently,
molecular formation barely affects the
densities of the neutral carbon and hydrogen that
determine the amount of CH (equation [\ref{cons_densidad1}]).
The CH abundance cannot depend very much 
on the actual set of molecules included when
solving equations (\ref{cons_densidad}) and (\ref{cons_densidad1}).
In order to find out which ones are really needed,
we solve them
including all atomic species (up to the second
ionization) and three different sets of molecules,
namely,
\begin{enumerate}
\item \label{case1} CH and H\( _{2} \),
\item \label{case2} CH, CO and H\( _{2} \),
\item \label{case3} CH, CO, H\( _{2} \), H\( _{2}^{+} \),
C\( _{2} \), N\( _{2} \), O\( _{2} \), CN, NH, NO, OH and H\( _{2} \)O,
\end{enumerate}
where the third case is supposed to represent the exact solution.
We calculate the abundances in  a
set of typical photospheric model atmospheres,
trying to cover all possible situations: from sunspot umbrae to network magnetic
concentrations, including the quiet Sun.
The general behavior  is very well illustrated by the
two examples  in Figure \ref{quiet2}. 
The number of CH molecules corresponding to the
cases \ref{case2} and  \ref{case3} differ by less than 0.01\% in the
quiet Sun. 
Even if the formation of CO is neglected (case \ref{case1}),
the errors are smaller than 10\% (see Figure \ref{quiet2}a). 
The approximations  corresponding to the
cases \ref{case1} and \ref{case2} work even better
in network and plage regions. 
Sunspot umbrae are different, though.
Here the CO exhausts the carbon available to create CH and
its formation must be considered
(Figure \ref{quiet2}b).
The 
approximation in case \ref{case2} still holds within 1\% .
Having in
mind these results, we calculate all our CH abundances
using just CO and H$_2$, thus being precise but
simplifying a lot 
the numerical solution of the
complete system of non-linear equations.

\placefigure{quiet2}

\subsection{Molecular lines\label{mlines}}

Once the abundance $n_{\rm CH}$ is obtained, one can 
evaluate the CH absorption.
Given an energy level in the molecule, 
its population follows from the Boltzmann equation,
\begin{equation}
\label{boltzmann}
n_{evJ}=n_{\rm CH} \frac{g_{eJ}}{U_{\rm CH}(T)} \exp [{-E_{evJ}/(kT)}],
\end{equation}
where the quantum numbers $e$, $v$ and $J$ denote
the electronic level, the vibrational level inside each electronic level, and
rotational level within each vibrational level, respectively. 
The other symbols have the usual meaning;
$g_{eJ}$ is the degeneracy (which only depends on the 
electronic and rotational quantum numbers), $E_{evJ}$ is the energy of the level, and 
$U_{\rm CH}(T)$ is the total partition function. 
%This partition function can be 
%calculated using
%\begin{equation}
%\label{partition}
%U_{\rm CH}(T)=\sum_{e,v,J} g_{eJ} \exp [{-E_{evJ}/(kT)}],
%\end{equation}
%where the summation extends to all molecular energy levels.
The line opacity for a transition from 
a level $e v J$ is
\begin{equation}
\label{lineopacity}
\chi_{i}=2.65\times 10^{-2} f_i
\Big(1-\exp[-h\nu_i /(kT)] \Big)\ n_{evJ},
\end{equation}
$f_i$ being the oscillator strength and $\nu_i$ the frequency 
of the transition. 
The constant above requires
the lower level population in cm$^{-3}$
to give $\chi_i$ in Hz cm$^{-1}$.
Due to the richness of the CH spectrum and the operation of
various line broadening mechanisms, many different transitions
contribute to the absorption coefficient at each frequency $\nu$,
namely,
\begin{equation}
\label{totalopacity}
\kappa_{\rm CH}(\nu) = 
	\sum_{i}{\frac{\chi_i}{\Delta \nu_D \sqrt{\pi}} 
H\Big(a,\frac{\nu-\nu_i}{\Delta \nu_D}\Big)}, 
	\label{line_shape}
\end{equation}
where the individual transitions are assumed to be
Voigt functions $H$ with a single damping parameter $a$ and a single
Doppler width $\Delta\nu_D$. The second assumption is well founded
since the thermal and
micro-turbulent motions produce almost the same broadening
across the \gband  . The damping is supposed to be constant 
for the sake of simplicity.  

We sum equation (\ref{totalopacity})
considering all CH lines in the spectral region between 4295 \AA\ and 4315 \AA\,
tabulated by
\citeN{jor96}. There are 746 transitions,
90 \% of them belonging to the 
$A^{2}\Delta - X^{2}\Pi$ electronic system.
The work by \citeN{jor96} also renders
the rest of parameters required to evaluate $\kappa_{\rm CH}$,
namely,
the weighted oscillator strengths $g_{eJ}\times f_i$, the frequencies $\nu_i$,
and the excitation potentials $E_{evJ}$.
As for the temperature dependence of the partition function,
we employ the approximate analytic expression given \citeN{sau84}.
We choose it
for consistency with the equilibrium constants used 
in \S \ref{cha}.
The damping $a$ is set by
comparison of synthetic spectra with the observed solar spectrum
(\S \ref{the_quiet}). In order to
carry out this comparison, the frequencies
have to be transformed to observed wavelengths. This
change 
involves the use of the speed of light in the
air $c_{\rm air}$,
\begin{equation}
\lambda = c_{\rm air}/\nu, 
\end{equation}
that we evaluate according to the formula by 
\citeN{edl66}.

%%%%%%%%%%%%%%%%%%%%%%%%
%
\subsection{Time dependent chemical modeling 
and CH photo-dissociation\label{photod}}

We use TE to estimate the CH abundance, which does not
imply any explicit assumption on the mechanism
responsible for the molecule formation.
In order to analyze which physical processes generate the
observed 
CH, we produced a Time Dependent
Chemical Model (TDCM) describing the formation of the
main molecules under typical photospheric conditions.
We applied a TDCM code developed within the framework
of a project on non-LTE radiative transfer 
in molecular lines. Details will be given elsewhere; here we just
brief those parts relevant for our simulations.

The chemical evolution of a system is described by the following
set of nonlinear ordinary differential equations (e.g., \citeNP{ben88}),
\begin{eqnarray}{}
\nonumber
\label{evolution}
\frac{dn_{m}}{dt}=
	\sum_{\rm A, B, C}\alpha_{\rm ABC}\ n_{\rm A}n_{\rm B}n_{\rm C}+
	\sum_{\rm A, B}\alpha_{\rm AB}\ n_{\rm A}n_{\rm B}+\sum_{\rm A}\alpha_{\rm A}\ n_{\rm A}\\
	-n_{m}\sum_{A, B}\alpha_{{\rm AB}m}\ n_{\rm A}n_{\rm B}
	-n_{m}\sum_{\rm A}\alpha_{{\rm A}m}\ n_{\rm A}-n_{m} \alpha_{m},
\end{eqnarray}
where $ n_{m} $ is the density of species $m$, and $m$ runs through all the species
included in the problem. The first term represents three-body reactions which generate
$m$ ($ {\rm A}+{\rm B}+{\rm C}\longrightarrow m $ + products), the second term corresponds to
two-body
reactions ($ {\rm A}+{\rm B}\longrightarrow  m$ + products), and the third
one describes one-body processes (${\rm A}\longrightarrow m$ + products). The negative
terms represent destruction of the molecule $m$ via three-body, two-body and
one-body reactions, respectively. Usually one-body reactions are 
photo-dissociation
or photo-ionizations, 
where the molecule is dissociated by photons of energy greater
than the dissociation energy, or it is ionized by photons of lower 
energy. 
The coefficients  $ \alpha$ stand for the reaction
constants, which depend on temperature and quantify the reaction rates
of the processes.
In order to get reliable results out of equation (\ref{evolution}),
it is desirable to
include all possible reactions which can produce or destroy the molecules.
Incompleteness in the reaction database will produce differences between TE
and TDCM results, even if TE is a good approximation. TDCM has been extensively
used in the interstellar medium (see, e.g., \citeNP{van98}),
typically including  only two-body reactions, photo-ionization
and photo-dissociation, as the set of fundamental processes. This approximation
is justified
due to the low densities that make three body reactions extremely rare.
The densities in the solar atmosphere favor three-body reactions, though.
For example, the catalytic molecular hydrogen formation reaction 
${\rm H} +{\rm H} +{\rm H} \longrightarrow {\rm H}_{2}+{\rm H}$
has a very small reaction rate but the density of H is so high
as to efficiently produce  H$_2$. 
There are some databases
with reactions, but they are oriented to interstellar chemistry (e.g. the UMIST
database, \citeNP{ben88}), and they do not include the full set of reactions 
needed for the Sun. In our case it
is more appropriate  to use a database of combustion reactions,
which have a better coverage of the solar physical conditions.

We aim at investigating the importance of photo-dissociation in
the solar photosphere. 
It could perhaps explain the absence of CH in magnetic concentrations and,
therefore, the origin 
of the \gband\ BP (see \S \ref{introduction}). In order to take it into account,
we need the photo-dissociation rates of the CH molecule
in the solar environment. CH photo-dissociation rates are computed for
the interstellar medium (\citeNP{van87}), where the ultraviolet radiation field 
responsible for the dissociation is much lower than in the Sun (\citeNP{dra78}). 
We will use them after a suitable
scaling.
The mean photo-dissociation rate is (e.g., \citeNP{rob81}), 
\begin{equation}
\alpha_{photo}=
\frac{4\pi}{hc}\int_{912 {\rm \AA}}^{3575{\rm \AA}}\lambda  J_\lambda\ \sigma(\lambda )d\lambda ,
\label{rates}
\end{equation}
where $ J_\lambda  $ stands for the mean intensity in the medium,
$\sigma(\lambda )$ corresponds to
the cross section for the photo-dissociation processes and, finally,   $c$ and $h$
have their usual meanings (speed of light and Planck constant, respectively). The integration 
extends
from the Lyman limit at 912 \AA\ (bluer photons 
are absorbed by the neutral hydrogen) to the photo-dissociation wavelength
at
3575 \AA. It can be formally extended to infinity, but the contribution 
of photons with energy lower than 3.47 eV is negligible since 
$\sigma(\lambda)$ falls off 
beyond this point. CH dissociates mainly through
bound levels of excited states that can couple with the continuum
of a dissociative state of the same symmetry
(\citeNP{van87}; for a discussion on photo-dissociation processes, see \citeNP{van87a}). 
The shape of the cross section in this case depends on details of the 
molecular
structure, but it always consists of a continuous background with superimposed
resonances. \citeN{kur87} tabulates the CH photo-dissociation
cross sections for a discrete set of temperatures describing the
population of the CH energy levels. We use them 
to integrate equation (\ref{rates}) under the assumption
that $J_\lambda$ is given by the local Planck function. The 
results for the highest and lowest photospheric temperatures are
given in Table \ref{table0}.
The table also contains
the mean lifetime of a CH molecule due to photo-dissociation,
$\overline{t}=1/\alpha_{photo}$. Some of these
lifetimes are fairly small ($2 \times 10^{-4}$ s),
indicating that it could be an efficient
process in the solar chemistry. However, 
these numbers are not enough to conclude whether photo-dissociation is important,
since one has to take into account all possible branches producing and
destroying CH. Such a simple molecule results from dissociation of
more complicated molecules, and can be easily generated in three-body reactions.

\placetable{table0}

\begin{figure}
	%\plotone{/net/mero/scratch/aasensio/textos/jorge/time_evol.ps}
	%\plotone{/home/jos/texto/papers/paper32/time_evol.ps}
	%\plotone{fig3.ps}
	\plotone{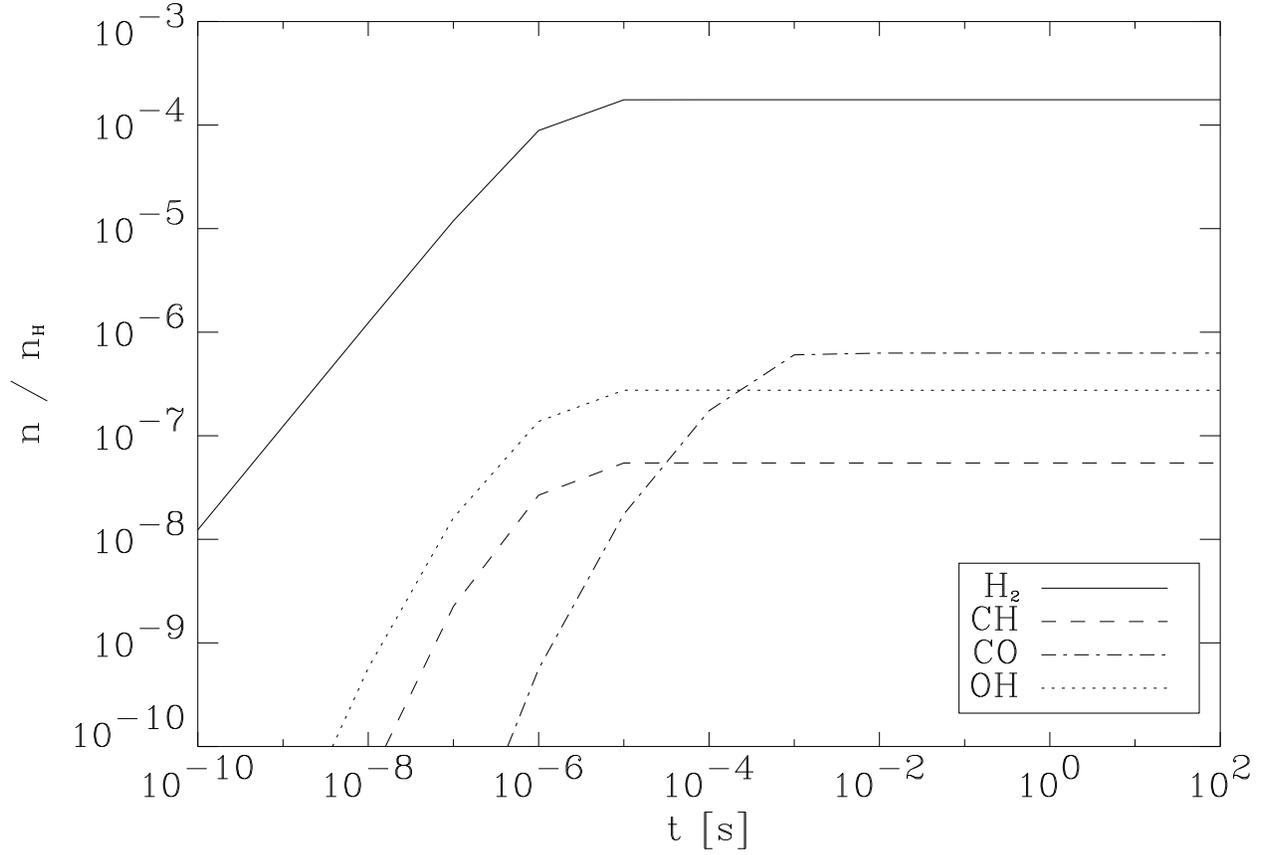}
\caption{Time evolution of the abundances of H$_2$, CH, CO, and OH 
(refer to the inset for the correspondence between type of line and
species). This integration of the chemical evolution equations
corresponds to typical photospheric conditions with no molecules at first
(T= 6500 K and initial $n_{\rm H}=1.7\times 10^{17}$ cm$^{-3}$).
The (thermodynamic) equilibrium CH  abundance
is reached within $10^{-5}$ s, a time-scale primarily set by the
reactions that
produce H$_2$ (see text and compare the solid and the dashed
lines). Photo-dissociation  seems 
to play a minor role in determining the CH abundance and so its
exact rate does not modify these curves.
All abundances, $n$, are referred to the abundance of atomic H, $n_{\rm H}$.
	\label{time_evol}}
\end{figure}

To  calculate the full TDCM,  we use a code that solves
the set of equations (\ref{evolution}) once the temperature and
density are given.
This set of equations represents a very stiff problem with the
different 
variables having disparate ranges of variation. A suitable method
which can cope with the stiffness has to be used; we employ 
an algorithm based on the backward differentiation formula to assure stability
(see, e.g., \citeNP{gea71}). As far as the reactions are 
concerned, we adopt a set of 110 neutral-neutral reactions
involving the following species:  
H, C, O, N, He, CH, CO, H$_2$, OH, NH, N$_2$, NO, O$_2$, HO$_2$, H$_2$O, 
and H$_2$O$_2$. Their reaction rates come from the
small hydrocarbons combustion 
database by \citeN{kon00}. Although reactions involving ions
should be included for comprehensiveness, our assumption
is reasonable
considering that
the main atoms involved in the CH formation (H, C, and O)
remain neutral under typical photospheric conditions.
The photo-dissociation rates come from Table \ref{table0}.
The solutions are initialized without molecules
for the hydrogen densities typical of photospheric
model atmospheres (see Table \ref{table0}). The 
rest of atomic species are taken according to standard solar 
abundances (\citeNP{gre84}). 
We have 
tested that the set of reactions
included in the TDCM code are adequate since the most important
species approach 
TE abundances when the solutions become stationary.
(Small differences and differences
in other minor molecules are to be expected because of the 
incompleteness of the 
database). 
Figure \ref{time_evol} shows the main molecular species in 
one of such simulations representing the bottom of the
photosphere.

\placefigure{time_evol}

Aided with the TDCM code, we
explore which processes produce the CH in the photosphere.
First we investigate the importance of photo-dissociation
as compared to the rest of the destructions mechanisms.
We define the parameter $r_{photo}$ as the  ratio
between the number of photo-dissociations and the total number of destructions
(see equation [\ref{evolution}]),
\begin{equation}
\label{ratio}
r_{photo}=\alpha_{photo}/\Bigl(\sum_{A, B}  \alpha_{ABm}\ n_{A}n_{B} + 
	\sum_{A} \alpha_{Am}\ n_{A} + \alpha_{m}\Bigr).
\end{equation}
This ratio in equilibrium is listed in Table \ref{table0} (actually,
we present
the ratio after 10$^2$ s).
Note that $r_{photo} < 2$ \%, even in the worst case scenario.
This clearly indicates that photo-dissociation
is of secondary importance for destroying CH in the 
Sun\footnote{The conclusion is
based on photo-dissociation
rates that assume the radiation field to be Planckian at the local
temperature. 
\modification{
We would have reached the 
same conclusion using a more realistic radiation field
to evaluate equation (\ref{rates}).
For example, even if $J_\lambda$ were ten times the local Planck
function, photo-dissociation still would be negligible.
}
}.
Actually, the analysis of the different destruction rates indicates
that, in all cases, CH is predominantly destroyed by collisions
with neutral hydrogen that yield molecular hydrogen, namely,
\begin{equation}
{\rm CH} + {\rm H}\longrightarrow {\rm C} + {\rm H}_2.
	\label{destruction}
\end{equation}
We have also inspected the creation rates, i.e., the
positive terms in equation  (\ref{evolution}). The CH
is produce via the reverse reaction (\ref{destruction}), i.e.,
\begin{equation}
{\rm C} + {\rm H}_2\longrightarrow {\rm CH} + {\rm H},
	\label{creation}
\end{equation}
the direct reaction between constituents
(${\rm C} + {\rm H}\longrightarrow {\rm CH}$) being some
two orders of magnitude slower.
The  tight link of CH to H$_2$ results clear
in Figure \ref{time_evol}, where the CH curve closely follows the 
curve 
of H$_2$. In particular, both reach the equilibrium within 
10$^{-5}$ s. This time scale depends very much on the densities
in the atmosphere, but it is extremely short and
always below $10^{-4}$ s.

\section{Local Thermodynamic Equilibrium G-band 
formation.\label{lteformation}}

\subsection{Atomic lines\label{lines}}

\begin{figure}
	%\plotone{/home/jos/texto/papers/paper32/figure2.ps}
%\plotone{fig4.ps}
\plotone{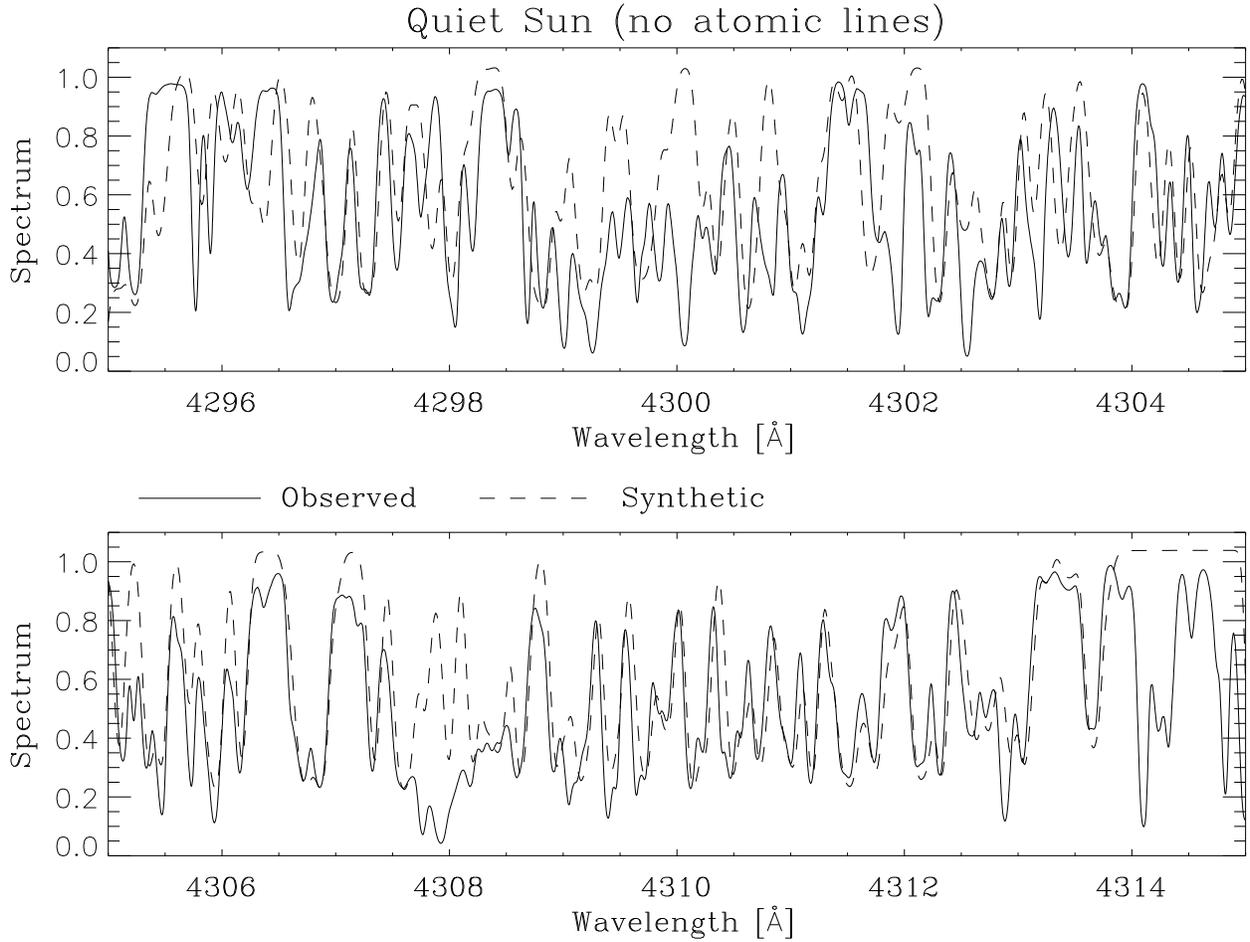}
\caption{Quiet Sun spectrum including the
	G-band modeled here. The solid line corresponds to
	the observed solar spectrum at the disk center
	(\citeNP{del73}). 
	The dashed line represents a best fit synthetic
	solar spectrum where the
	contribution of the atomic lines in the band
	pass has been neglected. Note the clear deficit
	of absorption at selected wavelengths. 
	Compare this fit with Figure \ref{quietsung},
	where atomic lines are included.
	Wavelengths are given in \AA , and the  specific intensity
	is normalized to the continuum intensity at the solar disk center.
\label{noatoms}
	}
\end{figure}

The \gband\ spectral region contains many 
atomic lines. Our first trial syntheses did not include them, however,
it 
became evident that they were needed 
for a proper reproduction of the observed spectral features.
The importance of the atomic lines can be judged
from inspection of Figure \ref{noatoms}. It
shows the \gband\ spectrum observed in the quiet Sun
together with a best-fitting synthetic spectrum 
without atomic lines. Every major
discrepancy can be attributed to the
contribution of at least one of such atomic lines. In order
to overcome this  deficiency,
we added the absorption produced by all relevant spectral lines 
in the region. Explicitly, we selected the atomic spectral lines
whose
quiet Sun disk center equivalent width is larger than
0.1 m\AA .  
This criterion renders some 70 lines in the 20 \AA\ interval 
between 4295 \AA\ and  4315 \AA . 
For convenience, the
search was carried out
using the database compiled by A. D. Wittmann, 
that includes 
excitation potentials and oscillation strengths
(see \S 2 in \citeNP{vel94}).
The absorption coefficients of the individual spectral
lines were computed from temperature and
electron pressure following standard procedures
(\citeNP{san97b}; \citeNP{san92a}). 
The abundances of the different elements come from \citeN{gre84},
whereas the selection of  the
damping is 
explained in \S \ref{the_quiet}.

\placefigure{noatoms}

\subsection{Spectral synthesis and \gband\ brightness\label{synthesis}}

The principles and rules detailed in \S \ref{ch} and \S \ref{lines} render
the absorption produced by the CH and the atomic lines.
By adding the H$^-$ continuum absorption, 
we calculate
the absorption coefficient $\kappa_\lambda$ that allows to integrate
the
radiative transfer equation,
\begin{equation}
{{d I_\lambda(z)}\over{dz}}=-\kappa_\lambda[I_\lambda(z)-B].
\label{rte}
\end{equation}
The symbols $I_\lambda$, $\lambda$, $B$ and $z$ stand for
the specific intensity, the wavelength,
the Planck function, and
the height in the atmosphere, respectively. 
Equation (\ref{rte}) is solved 2000 times per spectrum to
cover the range
between 4295 \AA\ and 4315 \AA\ with a 10 m\AA\ sampling interval.
The mean synthetic \gband\ signal,  $I_G$, 
results from 
direct numerical integration of the emerging spectrum, $I_\lambda(\infty)$,
\begin{equation}
I_G=\int_{4295 {\rm \AA}}^{4315 {\rm \AA}} f_G(\lambda) I_\lambda(\infty) 
d\lambda,
\label{ig}
\end{equation}
with  $f_G(\lambda)$ the color filter that describes
the \gband . For the sake of simplicity, we assume $f_G$ to be a 
Gaussian function whose central wavelength (4305 \AA) and
FWHM (12 \AA ) reproduce the device used by Berger and collaborators.
The band-pass of such filter is,
\begin{equation}
	f_G(\lambda)=f_0\exp\{-[(\lambda-4305\ {\rm \AA})/7.2\ {\rm \AA}]^2\},
	\label{gbandfilter}
\end{equation}
where the normalization constant $f_0$ ensures
\begin{equation}
\int_{4295 {\rm \AA}}^{4315 {\rm \AA}} f_G(\lambda) d\lambda=1.
\end{equation}
Figure \ref{quietsung} contains the filter band-pass
(equation [\ref{gbandfilter}] with $f_0=1$), together
with the \gband\ solar spectrum.

The use of equation (\ref{rte}) implies neglecting the polarization
of the light to compute the \gband\ spectrum. 
In view of the difficulties to compute the polarized spectrum
(see \citeNP{ill81}), the low degrees of polarization to be 
expected, and the exploratory nature of our calculations,
such assumption is well justified 
(see \citeNP{san99b}). The numerical integration of equation (\ref{rte})
has been completed using two different
techniques depending on whether we deal with plane parallel atmospheres
(short-characteristics, \citeNP{kun88}) 
or 
MIcro-Structured Magnetic Atmospheres 
(predictor corrector). The difference does not obey
technical reasons, rather it stems  from adapting
pre-existing codes to our purposes. 
Plane parallel atmospheres
are considered to represent the quiet Sun, intergranules, and the cores of the
large fluxtubes. 
MIcro-Structured Magnetic Atmospheres (\misma s)
describe  irregular magnetic concentrations having many assorted
optically-thin fluxtubes
overlapping along the line-of-sight.
\misma s offer a 
single framework that quantitatively reproduces 
the polarization produced
by plage, network and internetwork regions
(\citeNP{san97b}; \citeNP{san00}).
Despite the potential complexity of the underlying atmospheres,
the emerging  \misma\  spectrum can be computed as for a 1D atmosphere
by integration of 
the radiative transfer along a single ray. 
One has to replace in equation (\ref{rte})   
the absorption $\kappa_\lambda$ and the emission
$\kappa_\lambda B$
with 
the volume averaged absorption 
$\langle \kappa_\lambda \rangle$ and the volume averaged
emission $\langle \kappa_\lambda B\rangle$ (\citeNP{san96}; \citeNP{san97b}).
As for 1D atmospheres, the radiative
transfer equation admits a formal solution whose integrand
is  one of the so-called
contribution functions (hereinafter $CF$).
We will
write down  the solution since this $CF$ will
allow us to argue 
that the \gband\ signals are produced deep in the
photosphere\footnote{Beware
of simplistic interpretation of the contribution functions, though.
They are not unique (\citeNP{mag86}), and their use
to assign height-of-formations results 
ambiguous
(\citeNP{san96b}).} (\S \ref{hofs}). By defining
the wavelength mean contribution function  of the 
\gband\ as
\begin{equation}
CF(z)=\int_{4295{\rm \AA}}^{4315 {\rm \AA}} f_G(\lambda)\Bigl[
	\langle \kappa_\lambda B \rangle \exp({-\int_z^{\infty}\langle 
\kappa_\lambda \rangle}dz') 
	\Bigr]d\lambda,
	\label{cf1}
\end{equation}
then the \gband\ signal in equation (\ref{ig}) results,
\begin{equation}
I_G=\int_{-\infty}^{\infty}CF(z) dz.
	\label{cf2}
\end{equation}

	The LTE synthesis described in this section
was coded in IDL. We choose it  for convenience, but it turns 
out to be  fairly slow. For example,
the synthesis in a MISMA model magnetic concentration takes
some 25 minutes in an Ultra Spark 5 SUN workstation. This suffices
for the type of exploratory calculations that we carry out but
it prevents any massive application of the code. Thinking of such 
an application, we studied
a trivial way to speed up the procedure by using the
wavelength mean absorption coefficient to solve
equation (\ref{rte}). (The computation of $I_G$ would require a single representative
integration instead of the 2000 wavelengths that we employ.) 
We warn against 
this trick which  does not work in practice due to the
strong saturation of many lines in the band-pass.

\subsection{The quiet Sun synthetic spectrum\label{the_quiet}}

\begin{figure}
	%\plotone{/home/jos/texto/papers/paper32/figure0.ps}
	%\plotone{fig5.ps}
\plotone{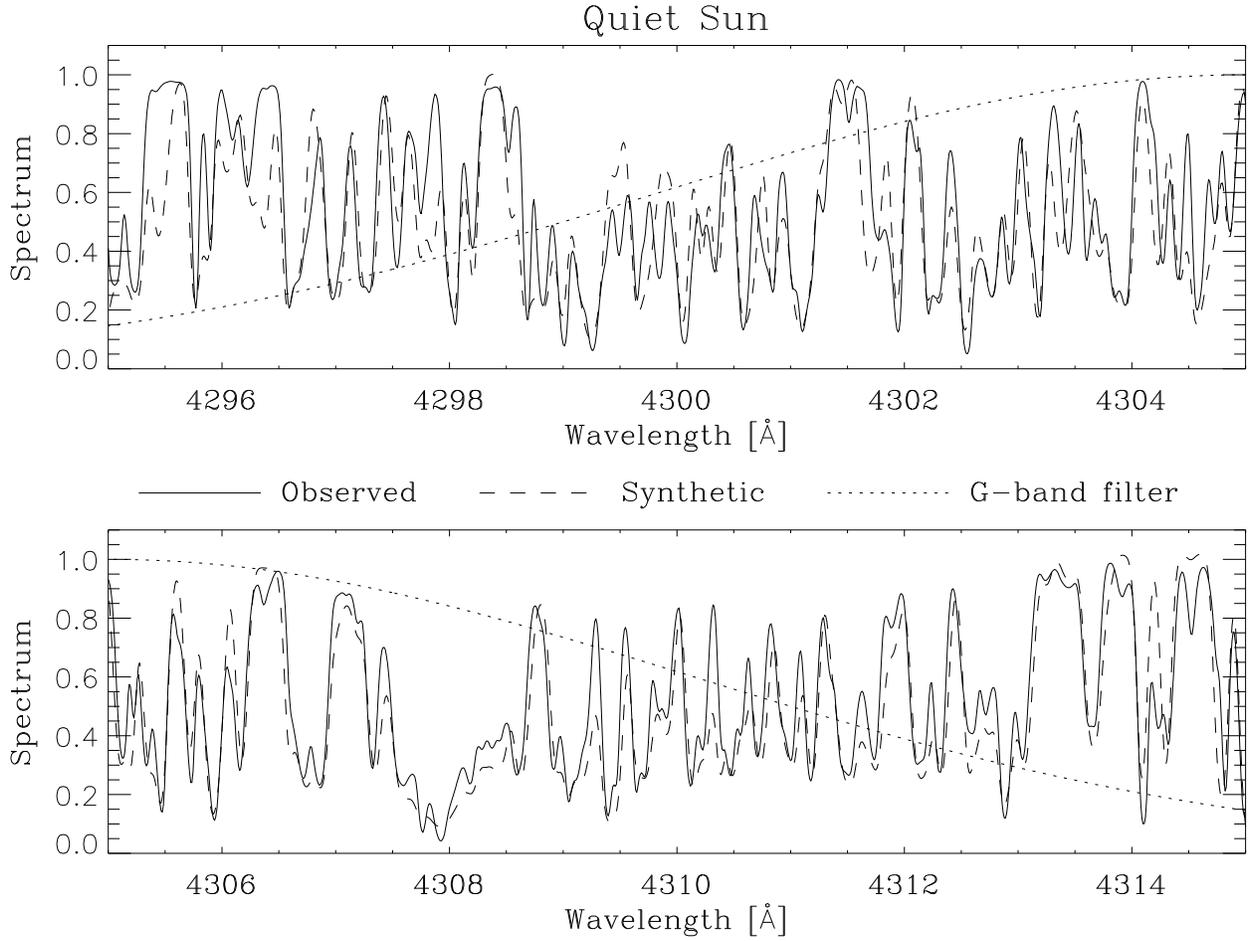}
\caption{Portion of the quiet Sun solar spectrum comprising the
	G-band as we model it here. The solid line corresponds to
	the observed solar spectrum at the disk center
	(\citeNP{del73}). The dashed line shows the synthetic
	solar spectrum obtained with our LTE code
	(\S \ref{the_quiet}). It is normalized to the observed continuum intensity.
	The dotted line corresponds to the
	Gaussian color 
	filter, 12 \AA\ FWHM, that we use to model observations 
	carried out in the G-band.
	Wavelengths are given in \AA .
\label{quietsung}
	}
\end{figure}

	Our \gband\ LTE synthesis code was tested 
against the observed solar spectrum.
We use the Li\`ege atlas (\citeNP{del73})
as reference observed spectrum, whereas 
the quiet Sun model atmosphere by \citeN{mal86}
represents the Sun.  The Li\`ege atlas 
is normalized to its local continuum, that we
transform to absolute units employing measurements of
the absolute solar continuum intensity at the
disk center
(\citeNP{all73}).
Figure \ref{quietsung} over-plots the observed and
the synthetic spectra. 
Considering the amount of information from
disparate sources that we have pieced together,
the degree of agreement
is remarkable. 
The difference of spectra has a standard deviation of
some 14\% (refereed to the continuum intensity). 
When the spectra are collapsed to get 
$I_G$ (equation [\ref{ig}]), then the
difference reduces to 6\%.  The discrepancies
are due to
localized 
deficits and excesses of opacity, e.g.,
overlooking of small molecular and atomic lines,
uncertainties in the line strengths and wavelengths,
departures from LTE, etc. Line
cores are well fitted, including the deep absorption at 4308 \AA\ produced
by the combined effect of Fe {\sc i}, Ti {\sc ii},
and CH. 
This overall good match, however,
required a tuning of the 
parameters that control the line shapes:
micro-turbulence and 
damping, according to equation (\ref{line_shape}), and
a final smearing with a Gaussian representing
macro-turbulence, spectral instrumental profile, 
and the like.
Even so, these are
only three free parameters that have to be compared with
the some 800 different transitions in the synthetic spectrum.
The selection of line broadening parameters
was carried out by trial and error. They  were first estimated
synthesizing only atomic lines. They are less numerous
than for CH lines, thus making the visual comparison with observations simpler. 
After this first approximation,
syntheses including
CH lines were carried out to find the final damping and turbulence.
The best fit was found for a damping of 0.02,  a
micro-turbulence of 0.9 km s$^{-1}$, and a macro-turbulence equivalent
to 3 km s$^{-1}$. Note that we use a single set of parameters for
all the lines in the band-pass, independently on whether they
are molecular lines or atomic lines.
Except for a few test calculations 
in \S \ref{ltevsobs}, the damping obtained in this way is
used throughout the text.

\placefigure{quietsung}

%The 6\% difference between 
%the observed and synthetic \gband\ brightness sets an upper
%limit to the
%uncertainties of our syntheses, since we 
%will always use ratios of synthetic \gband\ signals.
%An important part of the 6\% discrepancy arises
%from errors like
%missing spectral lines,
%erroneous abundances and oscillator strengths, deviations from LTE, etc.
%We expect that the influence of all these factors cancels
%(or at least gets strongly reduced) by using
%ratios of synthetic spectra. 

\section{Comparison with observations\label{comparison}}

	The code described in the
previous sections was aimed at comparing LTE syntheses
in model magnetic concentrations with 
G-band observations.  A fair agreement 
would support the goodness of the unpolarized 
LTE approximation to modeling \gband ~spectra.

\subsection{Summary of \gband\ observations\label{observations}}
This section compiles
observational results
that will be discussed in evaluating the 
goodness of the syntheses.
The observations correspond to
the disk center.
Brightness are referred to the mean quiet Sun and, 
by definition, the term {\it contrast} denotes
the brightness minus one (i.e., the brightness minus the
quiet Sun brightness,  
set to one by our normalization).
The results have been distilled from different works:
$+$ \citeN{kou77}; 
$\times$ \citeN{deb92};
$*$ \citeN{ber95}; 
$\Box$ \citeN{ber96}; 
$\otimes$ \citeN{tit96}.
$\bullet$ \citeN{ber98b}. The
symbols accompanying these citations  are used for quick reference
in the list of results given next.

	\begin{enumerate}
		\item\label{item1} The contrast in the G-band
		is about twice as large as the
		white-light contrast
		(broad-band imaging including lines and
		continuum)$*$. In quantitative terms,
		\begin{equation}
		I_G-1=1.91 (I_{WL}-1)+0.23,
		\label{law}
		\end{equation}
		where $I_G$ and $I_{WL}$ correspond to the
		\gband\  brightness and the white-light brightness,
		respectively $\bullet$.
		Individual BPs  present considerable scatter with respect to 
this 
		trend, though $*$. 
		\item The mean observed contrast (i.e., the mean $I_G-1$)
		is 0.31, with a standard deviation of  0.11. The
		extreme values range from  $-0.19$ to
		0.75 $*$. \label{contrast}
		\item\label{unresolved}	Many observed BPs are 
			not spatially resolved $\Box$ $\otimes$.
			The same happens
			when the BPs are imaged in 
			white-light $\times$, and continuum $+$.
		\item The
		\gband~  brightness does not depend
		on the size 
		of the BP $*$.\label{nosize}
		\item There is also a diffuse component of
		G-band bright structures $\bullet$.
		\label{diffuse}
		\item \label{law2} The \gband~ contrast averaged
		over a magnetic network region is positive,
		whereas it is zero in white-light
		images $\bullet$.

		\item \label{granulation} 
		\gband\ and white-light images both show the solar granulation with
		identical appearance $\bullet$. The difference
		of these two types of images is used for automatic identification of the
		\gband\ BPs $\bullet$. 
	\end{enumerate}

\subsection{LTE syntheses versus observations\label{ltevsobs}}

\begin{figure}
	%\plotone{/home/jos/texto/papers/paper32/figure3.ps}
	%\plotone{fig6.ps}
\plotone{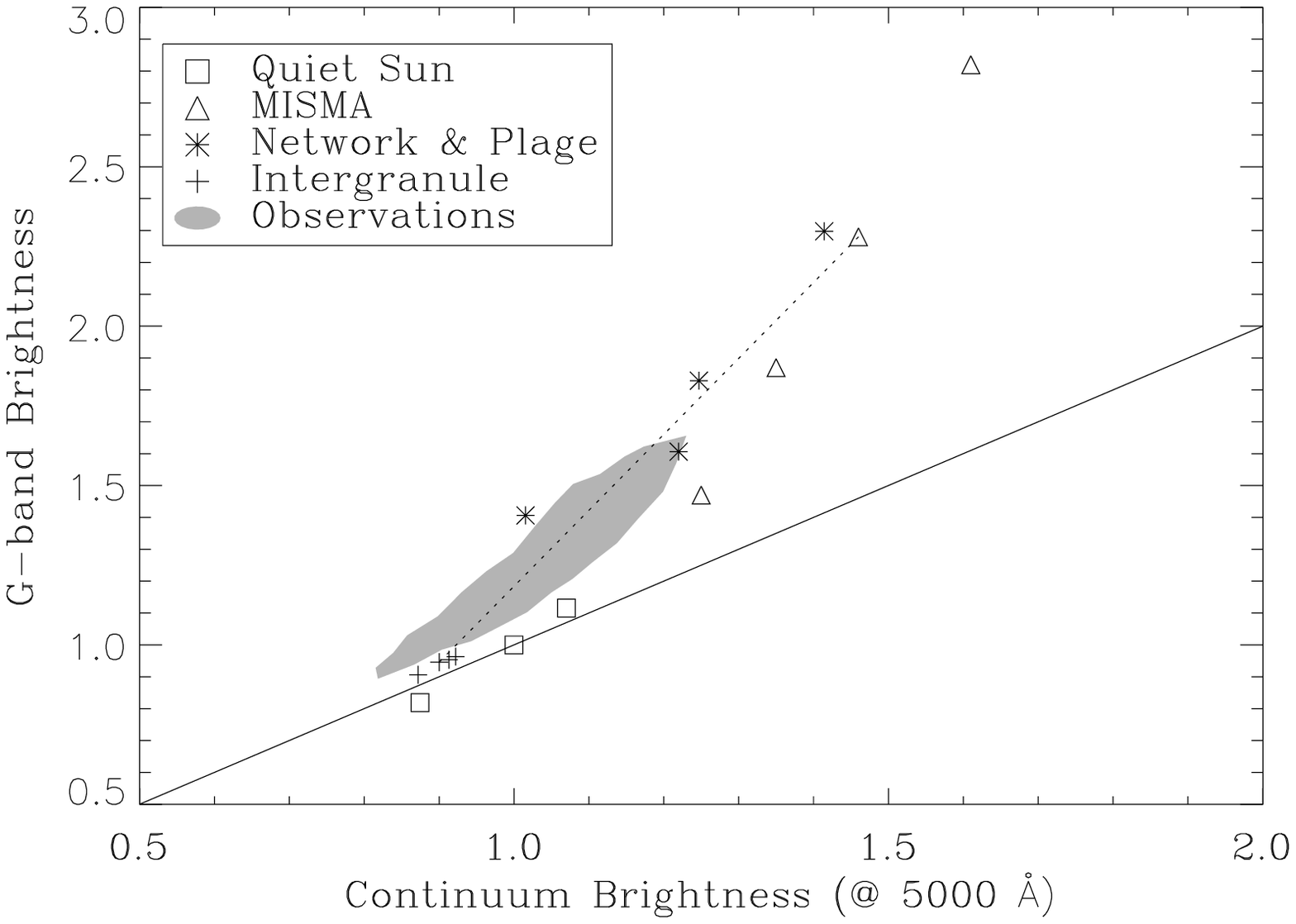}
\caption{LTE syntheses of the \gband\
	brightness produced by different
	model atmospheres at the disk center. 
	The stars $*$ represent network and plage model
	atmospheres based on
	single fluxtubes (\citeNP{cha79};
	\citeNP{sol86}; \citeNP{bel98b}). The
	triangles $\triangle$ correspond to model \misma s (\citeNP{san00}).
	Quiet Sun atmospheres are shown as squares $\Box$
	(\citeNP{gin71}; \citeNP{ver81}; \citeNP{mal86}).
	The plus signs $+$ describe the intergranular
	environment of the \gband~ bright points. 
	The dotted line shows all possible locations of a  magnetic
	concentration that is not spatially resolved. It has to be
	on the straight line that joins the
	fully resolved feature and the background; the actual
	position along this line is set by the filling factor.
	Observations by \citeN{ber98b} are shown for reference
	(the shaded area). 
	Atmospheres with 
	equal contrast in the \gband~ and 
	continuum have to lie on the solid line. 
	Everything has been normalized to the
	\citeauthor{mal86} quiet Sun.
	\label{summary}}
\end{figure}

Figure \ref{summary}  shows synthetic  G-band brightness 
versus (5000 \AA) continuum intensity 
for model atmospheres of various kinds located at the solar disk center.
It includes network and plage model
atmospheres based on
single fluxtubes\footnote{We just synthesize the
ray along the axis of the tube.} 
(\citeNP{cha79};
\citeNP{sol86}; \citeNP{bel98b}),
model \misma s (\citeNP{san00}),
quiet Sun atmospheres 
(\citeNP{gin71}; \citeNP{ver81}; \citeNP{mal86}),
and intergranules
(modeled using  
stray light components of \misma s; see \citeNP{san00}).
The latter are
relevant since they represent
the  environment of the \gband~ BPs.
All intensities are normalized to the 	
quiet Sun spectrum produced by the \citeN{mal86} atmosphere, that
is very similar to the spectrum  observed in the
quiet Sun (see \S \ref{the_quiet}).
Figure \ref{summary} also includes \gband\ 
observations with the best angular resolution available up to date
(Figure 3 in \citeNP{ber98b}). The white-light
channel of these observations does not correspond to true continuum, though.
The consequences of comparing white-light 
with continuum will be analyzed later.

\placefigure{summary}

First note that the \gband\ contrast in magnetic concentrations
is about twice the 
continuum contrast. 
(For reference, the slope of the dotted line in 
Figure \ref{summary} amounts
to 2.3.) This ratio of contrasts is in good  agreement
with observations (item \# \ref{item1} in \S \ref{observations}).
There is, however, a serious discrepancy as far as the
actual contrast is concerned. The synthetic
spectra are much brighter than the observed BPs
(item \# \ref{contrast} in \S \ref{observations}).
This discrepancy occurs in both white-light and
G-band and can be understood if even the
extremely high resolution data in which the observations are based
do not yet resolve the BP. The hypothesis is reasonable since
it stems from observations;
point \# \ref{unresolved}
in \S \ref{observations}. 
Assume this insufficient resolution.
The observed brightness $I_G$
has contribution from both magnetic structures $I_{G}^{m}$
and intergranular background $I_{G}^{b}$. Then,
\begin{equation}
I_G=\xi I_{G}^{m}+(1-\xi)I_{G}^{b},
\label{two1}
\end{equation}
being $\xi$ the fraction of signal due to the
magnetic concentrations. Similarly, the continuum intensity $I_c$ has also
contribution from the BPs $I_{c}^{m}$ and the
background $I_{c}^{b}$,
\begin{equation}
I_c=\xi I_{c}^{m}+(1-\xi)I_{c}^{b}.
\label{two2}
\end{equation}
According to equations (\ref{two1}) and (\ref{two2}), 
a spatially unresolved feature will show up in Figure \ref{summary}
on the straight segment that joins the points corresponding to the
fully resolved feature and the background.
The exact position
depends on $\xi$ which, in geometrical terms,  corresponds
to the fraction of this segment where the unresolved BP shows up.
Figure \ref{summary} displays one of such straight lines, specifically,
the one for a \misma\  and its intergranular background. It matches the
range of observed contrasts for $\xi\leq 0.5$.  The ability
to account for the observed contrasts is shared by all model
magnetic concentrations.
All magnetic models in Figure
\ref{summary} would  cross the shaded  area
for reasonable values
of the filling factor. This argument
can be put forth in quantitative terms
by combining
equations
(\ref{two1}) and (\ref{two2})
to form,
\begin{displaymath}
I_G-1=m~(I_c-1)+c,
\end{displaymath}
\begin{equation}
m=(I_{G}^{m}-I_{G}^{b})/(I_{c}^{m}-I_{c}^{b}),
\label{slope}
\end{equation}
\begin{displaymath}
c=m~(1-I_{c}^{b})-(1-I_{G}^{b}).
\end{displaymath}
As we pointed out above, \misma s and single fluxtubes all yield,
\begin{equation}
m\approx 2.
\label{two3}
\end{equation}
A glance
at the intergranular contrasts in Figure \ref{summary} indicates that
$(1-I_{c}^{b})\sim (1-I_{G}^{b})\sim 0.1$, consequently, equations (\ref{slope}) 
and (\ref{two3}) render
$c\sim 0.1$. The actual value is very uncertain but 
it has to be a small positive 
quantity,
i.e.,
\begin{equation}
0 < c \ll 1,
\label{two4}
\end{equation}
since 
$c=m~(1-I_{c}^{b})-(1-I_{G}^{b})\sim 
(m-1)(1-I_G^b)\sim (1-I_G^b)\ll 1$. 
The two results summarized in equations (\ref{two3}) and
(\ref{two4})
compare well with the observed law
(\ref{law}), and this agreement does not depend on which model magnetic
concentration is considered. (Obviously, we have assumed $I_{WL}=I_c$. The 
differences between the two quantities are analyzed below.)

Equations (\ref{slope}) and  (\ref{two4}) 
can also explain the positive mean \gband\ contrast
of magnetic regions having neutral continuum
contrast (point \# \ref{law2} in \S \ref{observations}). Assuming
$m$ and $c$ to be fairly constant across the magnetic
region, then
the relationship (\ref{slope}) applies to the 
mean quantities. 
Equations (\ref{slope})  and  (\ref{two4}) 
predict $I_G=1+c > 1$ when $I_c=1$.

\begin{figure}
	%\plotone{/home/jos/texto/papers/paper32/figure4.ps}
	%\plotone{fig7.ps}
\plotone{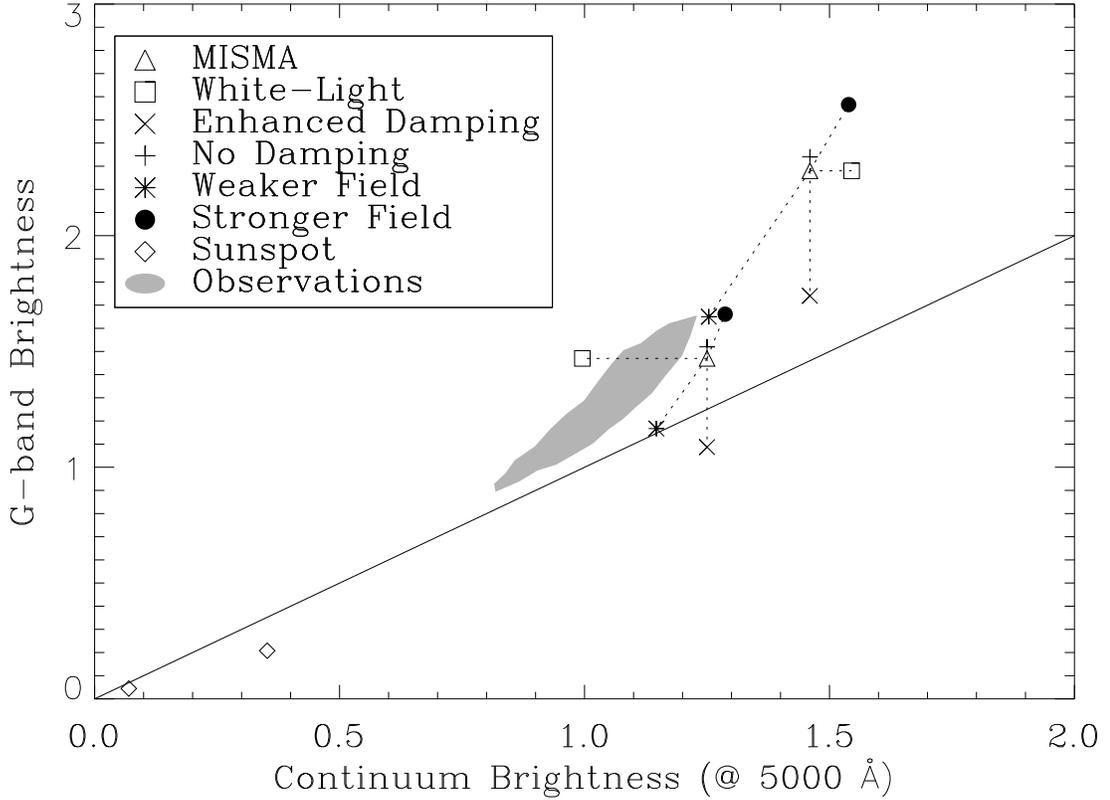}
\caption{
	Displacements of the \gband\ brightness and continuum brightness
	caused by modifications of the model atmospheres.
	\modification{
	Two model MISMAs illustrate the common behavior.
	The original points are shown as triangles $\triangle$.
	The dotted lines join them with the
	various transformations. 
	Including spectral lines in the continuum channel
	may increase or decrease the contrast; the
	continuum brightness of the brighter atmospheres
	become even larger
	 and vice versa (squares $\Box$). 
	Increasing the line wings reduces the \gband\ signals 
	(times signs $\times$). Setting the 
	damping to zero produces 
	the plus signs $+$.
	Decreasing the magnetic field strength makes everything darker
	(stars $*$). The opposite happens when the field
	increases (black dots $\bullet$).
	Note that the brighter model atmosphere remains bright
	under all circumstances
	whereas the darker atmosphere sometimes does not.
	The diamonds correspond to sunspot umbrae.
	}
	For the rest of the symbols, see
	Figure \ref{summary}.
	\label{cases}}
\end{figure}

\placefigure{cases}

So far we have considered a few representative model magnetic 
concentrations. 
In addition to these syntheses, we carried out a set of simulations
to find out 
how the synthetic contrasts in Figure \ref{summary} depend on details
of the atmospheres. We used \misma\  model atmospheres
with the claim that their behavior is  representative
of the other atmospheres.
First, we
explored how the position in the figure depends on using 
white-light instead of continuum intensity (except for the work by
\citeNP{kou77}, the white-light channel of the observations 
in \S \ref{observations}
includes many spectral lines).
We took a shortcut by replacing the true white-light
band-pass  with our \gband\ spectrum without CH lines
(the \gband\  synthesized considering only
continuum and atomic lines).
Two arguments justify the approximation. First, we just aim at a qualitative
description.  Second, the some 70 atomic
lines included in our \gband\ modeling 
should be representative of a typical
band-pass in the solar spectrum.  Two results 
are shown in Figure \ref{cases}. The original points,
triangles, are to be compared with the new points, boxes.
Including spectral lines in the band-pass induces two competing 
effects. The
lines weaken with increasing temperature, which
increases the white-light contrast with respect to the continuum
contrast. 
On the contrary, the turbulence
is enhanced in the magnetic atmospheres as compared with the
quiet Sun (e.g., \citeNP{sol93};
\citeNP{san00}). This enhances the absorption of the saturated
lines with respect to their absorption in the  quiet Sun, decreasing the
contrast in white-light.
The first effect dominates in the brighter (hotter) atmospheres, where
the lines are weak. The decrease of white-light contrast triumphs in  the
cooler atmospheres, though. Which tendency takes over depends on details of the
atmosphere (in particular on the micro-turbulence).

Another source of error of the \gband\ synthesis
is the damping wings of the lines, a very uncertain parameter. 
In the reference simulations of
Figure \ref{summary} we borrowed the damping
from the fit to the quiet Sun spectrum (\S \ref{the_quiet}). The
damping in the quiet Sun is probably not the same as in magnetic concentrations
and,
since we are dealing with many saturated lines in the band-pass,
moderate differences may play a significant role in determining
the \gband\ absorption.
To study the influence
of the wings, we synthesized \gband\ spectra  
using a damping 0.5, instead of the 0.02 
derived from the quiet Sun.
The G-band brightness decreases as expected due to
the global increase of absorption in the band-pass (see
the times signs in Figure \ref{cases}).
This decrease dims so much the \gband\ signal of
the cooler atmosphere in Figure \ref{cases} that it becomes even darker than
the quiet Sun. 
\modification{Figure \ref{cases} 
	includes the case with no
	damping for reference (the plus signs). 
}

\modification{
We have also varied
the magnetic field strength
of the original model MISMAs.
}
Note that the influence of this parameter on the
\gband\ brightness is not due to Zeeman effect,
that we neglect (\S \ref{synthesis}).
The modifications result  from the strong
coupling of the thermodynamic  parameters to the 
magnetic field strength. In particular, a reduction
of field strength implies an increase of gas pressure
to maintain the mechanical balance of the atmosphere
(e.g., \citeNP{san00}). An increase of gas pressure
produces higher densities so the atmosphere is more
opaque and the spectrum is formed higher up at lower temperatures.
\modification{ 
Consequently,
the decrease of field strength lowers the
contrasts.
A 20\% reduction 
is represented in Figure \ref{cases} using stars $*$.
The augment of
field strength leads to the opposite effect. 
We increase it up to  98 \% of the maximum possible
value, which induces the enhancement of contrast shown in
Figure \ref{cases} with black dots $\bullet$. 
Such large field strength almost empties the magnetic components
of the
MISMAs, so that the spectra emerge from
the non-magnetic components (see \citeNP{san00c}).
}

\modification{
A simple conclusion follows from the
exploratory  simulations described above:
the exact \gband\ synthetic contrast depends on details 
of the modeling difficult to constraint.
If a model BP is bright enough then such details
are secondary since the BP remains 
bright both in white-light and the \gband . However,
such variations can modify
the character of the
concentrations that are not so bright.
In particular, 
the relationship between continuum and \gband\ is expected to 
unsharpen or even
break down for intrinsically low contrasts, where continuum
BPs may or may not be \gband\ BPs depending on details of the atmosphere.
In keeping with this conclusion, the fair correlation that we observe
for weak contrasts is probably due to the limited 
angular resolution.
It would be produced by
intrinsically bright magnetic features mixed up
with large amounts of dark background
(as described by equation [\ref{slope}] with $I_c \gtrsim 1$).
On the other hand,
the expected fuzzy relationship between low contrast \gband\ BPs and 
contimuum BPs 
may be related to
the diffuse \gband\ component (item \# \ref{diffuse} in \S \ref{observations}).
The \gband\ signals would 
originate in  concentrations whose properties
lay on this ridge where to be or not to be
bright depends on details. 
}

\modification{
Figure \ref{cases} includes the synthetic contrast produced
by two model
sunspot umbrae: 
the very cold model M in \citeN{mal86}, and a hotter
one (\citeNP{col94}).
Unlike the quiet Sun magnetic concentrations studied up to now,
the umbra \gband\ contrast is lower than the continuum 
contrast.
}

%The conclusion to be extracted  from all these exploratory 
%simulations is simple:
%the exact \gband\ synthetic contrast depends on details 
%of the modeling difficult to constraint.
%However,
%should a model BP be bright enough then the details
%are secondary since the BP remains 
%bright both in white-light and the \gband . On the
%contrary, the character of \gband\ BP for 
%magnetic concentrations that are not so bright may be modified upon 
%variations
%of the atmospheric properties.
%In keeping with this conclusion, the fact the \gband\ observations correlate
%with white-light, even for weak contrasts, has to follow from the lack of
%angular resolution rather than from a true correlation. One should expect
%that the relationship breaks down for small contrasts, where continuum
%BPs may or may not be \gband\ BPs depending on details of the atmosphere.
%As a matter of speculation, let us mention that
%the diffuse \gband\ component (item \# \ref{diffuse} in \S \ref{observations}) 
%may be produced by concentrations whose properties
%lay on this  ridge where to be or not to be
%bright depends on subtleties.
%
%
%

\section{Corollaries of the LTE \gband\ modeling}

\subsection{Contribution functions for the \gband\ signals\label{hofs}}

	The contribution functions 
	($CF$s) in equations (\ref{cf1}) and (\ref{cf2})  
describe the atmospheric heights  producing synthetic signals. 
Figure \ref{cfs} shows continuum and \gband\ $CF$s 
in both a magnetic concentration 
and its intergranular environment. These $CF$s are normalized
to render the signals in units of the quiet Sun signals. Moreover,
the scale of heights is referred  to the 
optical depth unity in the quiet Sun. The functions
in 
Figure \ref{cfs} illustrate the
general behavior.

\placefigure{cfs}

Note first that the \gband\ and the continuum $CF$s 
are similar so that they peak at the same heights (compare
continuum and \gband\ $CF$s formed in the same atmospheres).
Second, all $CF$s reach a maximum below the zero of the scale of heights.
It is well below zero in the magnetic 
concentration because of the evacuation forced by the
presence of a magnetic field (the so-called Wilson depression).
On the other hand, the maxima of the intergranule $CF$s are very close 
to the quiet Sun  $CF$s (that we do not represent in the figure). 
Problems of interpretation  
notwithstanding, the analogy between continuum and
\gband\ $CF$s evidences that the corresponding signals
come from neighboring atmospheric layers. This result agrees
 with observations.
\gband\ and white-light images both show the solar granulation with
identical  appearance (see point \# \ref{granulation} in \S 
	\ref{observations}).
\citeN{rut99} points out that such similarity implies
\gband\ signals being formed deep in the photosphere where the
continuum is formed. We endorse his conclusion
which should be regarded as an observational constraint.
In our particular case, such observational requirements
is fulfilled in a natural way.

The obvious difference between  \gband\ and continuum $CF$s
lies in their different widths, the \gband\ $CF$s being broader (Figure \ref{cfs}).
This extended contribution follows from the varied range of opacities 
producing
the \gband\ spectrum; from the cores of strong lines to 
the continuum.

\subsection{Why are model magnetic concentrations so conspicuous
	in the G-band?\label{why}}

	We found that the LTE spectrum 
in almost any model magnetic concentration explains the observed 
G-band enhancement (\S \ref{ltevsobs}). 
\modification{
This section sketches the radiative transfer effect that
we believe to be responsible for such general behavior.
}

	\placefigure{physics}

	Figure \ref{physics} shows the variation with temperature 
of the absorption coefficients that are relevant for the LTE synthesis
of the G-band and the continuum.   
\modification{
It shows wavelength mean absorption coefficients at a constant
density. 
The wavelength mean weights the absorption with the color filter in
equation (\ref{gbandfilter}), whereas the density has been set 
to a value typical of the bottom of the quiet Sun photosphere.
The actual filter and density do not modify
the forthcoming qualitative arguments, though. A different filter
barely changes the mean absorption. A different density 
shifts all the curves in the vertical direction 
(they approximately scale with the density
within the range of temperatures represented in Figure \ref{physics}).
}
%(The wavelength mean weights the absorption with the color filter in
%equation [\ref{gbandfilter}], whereas the density has been set 
%to a value typical of the bottom of the quiet Sun photosphere.
%The actual filter and density do not modify
%the forthcoming arguments, though.)
Note how the opacity in the G-band has a minimum at some 6000 K 
which, roughly speaking, represents the temperature of the layers 
that are observed in the quiet Sun.
On the contrary, the continuum absorption coefficient does 
increase very rapidly with temperature. This different behavior
explains why the model magnetic concentrations are much 
brighter in the G-band than in 
continuum. All the model magnetic concentrations are hotter 
than the quiet Sun in 
the layers producing the observed light. These larger temperatures
enhance the continuum opacity, therefore, force the continuum photons 
to emerge from higher and therefore cooler layers.
The net effect is a damping of the radiation field 
increase that would 
correspond to the sole augment of temperature.
On the contrary, the opacity is barely
modified in the G-band (it may even decrease).
The temperature increase of the magnetic concentrations
directly 
augments the radiation field that escapes from the atmosphere,
enhancing the contrast of the magnetic concentrations 
when imaged in the \gband .

	The neutral dependence of the \gband\ opacity  on temperature
is produced by the fortuitous opposite behavior of the continuum H$^-$ opacity 
and the CH absorption. These two sources of opacity
determine the net effect.
The increase of H$^-$ opacity with temperature is due to the increase of free electrons
in the medium. The fall off of the CH opacity  just follows
the decrease of the CH abundance according to the Boltzmann factor
in the dissociation equilibrium (equations
[\ref{cons_densidad1}] and
[\ref{Equilibrium_cte}]),
\begin{equation}
n_{\rm CH} \propto \exp[D_0/(kT)].
\end{equation}
This dependence on the dissociation energy
becomes clear upon inspection of Figure \ref{physics},
where we include the approximate adsorption coefficient
resulting from the sole action of $D_0$, namely,
\begin{equation}
\kappa_{CH}\simeq \kappa_{CH}(T_0) \exp[D_0/(kT)-D_0/(kT_0)], 
\end{equation}
with $T_0=6000$ K. It reproduces the behavior of the mean opacity
in the region of interest. The discrepancy towards
low temperatures (say below 5000 K) is due to the formation
of CO molecules (see again Figure \ref{physics}, the dotted line). 
The CO formation  does not play a major role in our case, although it becomes
critical in pores and sunspots (see \S \ref{cha}).

To sum up, magnetic concentrations are so bright in the \gband\ because
they are hot. 
The increase of temperature with respect to the quiet
Sun is partly damped in the continuum,
where the opacity also increases with temperature.
The total \gband\ opacity barely depends on temperature, though.

%
%%%%%%%%%%%%%%%%%%%%%%%%%%%%
%
%
\section{Discussion and conclusions\label{conclusion}}

Bright Points
(BPs) in \gband\ images are associated with magnetic
concentrations, although the physical cause 
of such brightness has been debated
over the last years (see \S \ref{introduction}). 
The difficulties of understanding bear on the
complications of modeling a molecular band,
rather than on an enigmatic or unexpected 
behavior of the \gband\ signals. In order to 
explore
the simplest possibility,
we write and test a \gband\ LTE synthesis code
that explicitly includes all the CH lines
in the band-pass.  As it was conjectured
by \citeN{san00}, such simple LTE modeling
seems to account for the 
main observational facts. Specifically,
the \gband\ emerging from standard model magnetic
concentrations is very bright relative to the quiet Sun. The
\gband\ contrast
is about twice the contrast of the same structures
observed at continuum wavelength. 
The
synthetic contrasts, however, exceed by far the observed contrasts.
One can readily interpret this apparent contradiction
if
the observed \gband\ BPs are not yet spatially resolved. They
have to fill
50\% of the resolution  elements or even less.
Such requirement to match LTE syntheses and observations
holds for all the semi-empirical model magnetic
concentration that are available (from large fluxtubes
to \misma s; \S \ref{ltevsobs}). 
The lack of enough resolution is not an ad-hoc
assumption to fix up a conflict
with observations; it actually comes out of
the observations themselves (see \S \ref{observations}).
Insufficient resolution may also explain why the \gband\ contrast
does not depend on the size of seemingly resolved  \gband\ structures
(\S \ref{observations}).
Each observed \gband\ BP would be a conglomerate of
unresolved structures so that the 
actual observed contrast depends the
degree of mixing with dark non-magnetic surroundings, rather
than on the intrinsic contrast of the magnetic structures.
\modification{
Another possibility is the existence of a 
relationship between brightness and size masked
because the inadequate resolution. 
Small bright features and large faint ones show up with the same
apparent contrast 
(e.g., \citeNP{tit96}). 
}

\begin{figure}
%\figurenum{Caption}
	%\plotone{/home/jos/texto/papers/paper32/figure5.ps}
	%\plotone{fig8.ps}
\plotone{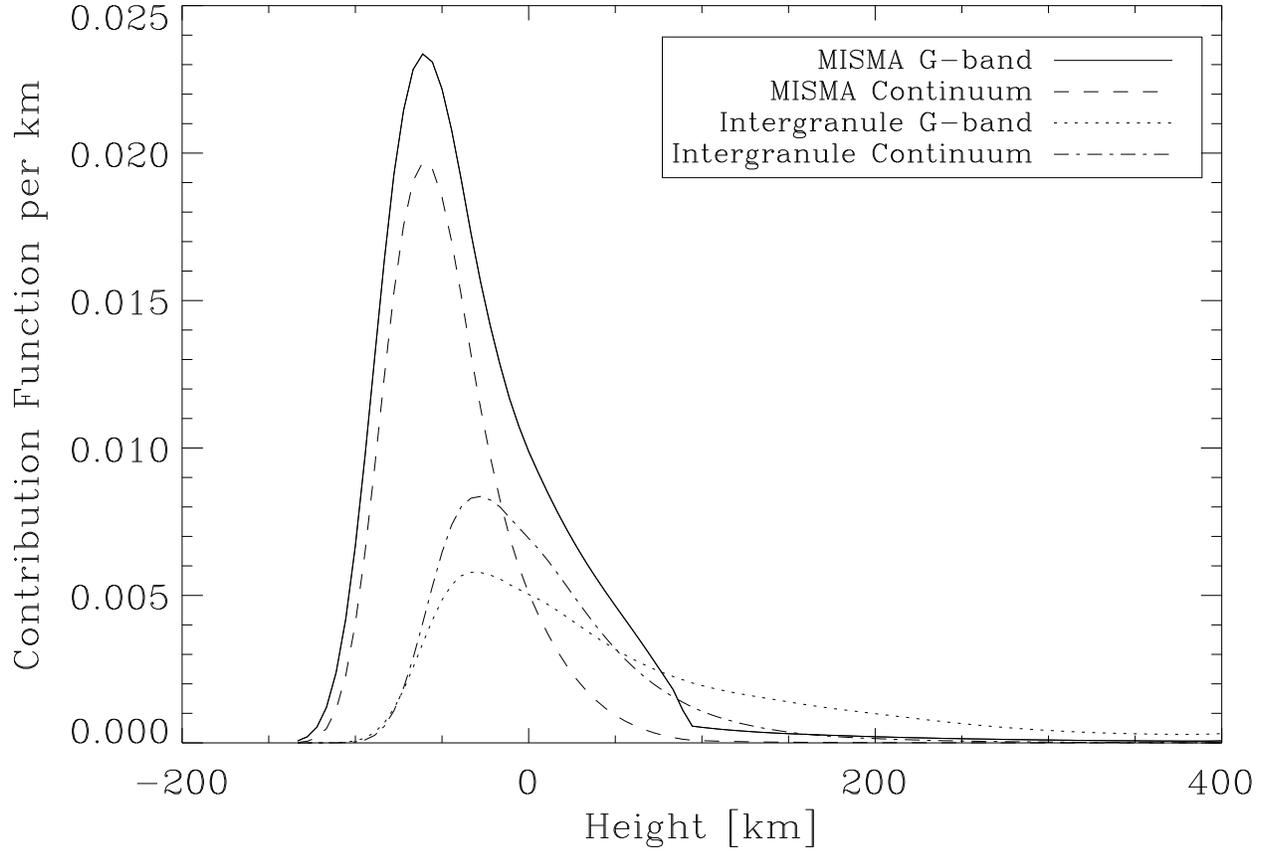}
\caption{
	Contribution Functions of the \gband~ and
	the 5000 \AA~continuum 
	in a typical model \misma . The $CF$s in the intergranular
	environment of the magnetic concentration are also included for comparison. 
	Note that the G-band $CF$ in the \misma\  peaks at some -60 km, which is
	very close
	to the maximum of the continuum $CF$.
	The $CF$s have been normalized so that their areas
	yield the intensities referred to the
	quiet Sun (\citeauthor{mal86} model atmosphere).  
	The inset on the upper right 
	corner provides the equivalence between type of line and
	$CF$. The 
	heights are referred to the continuum optical depth equals one
	in the quiet Sun. 
	\label{cfs}}
\end{figure}

\begin{figure}
	%\plotone{/home/jos/texto/papers/paper32/figure1.ps}
	%\plotone{fig9.ps}
\plotone{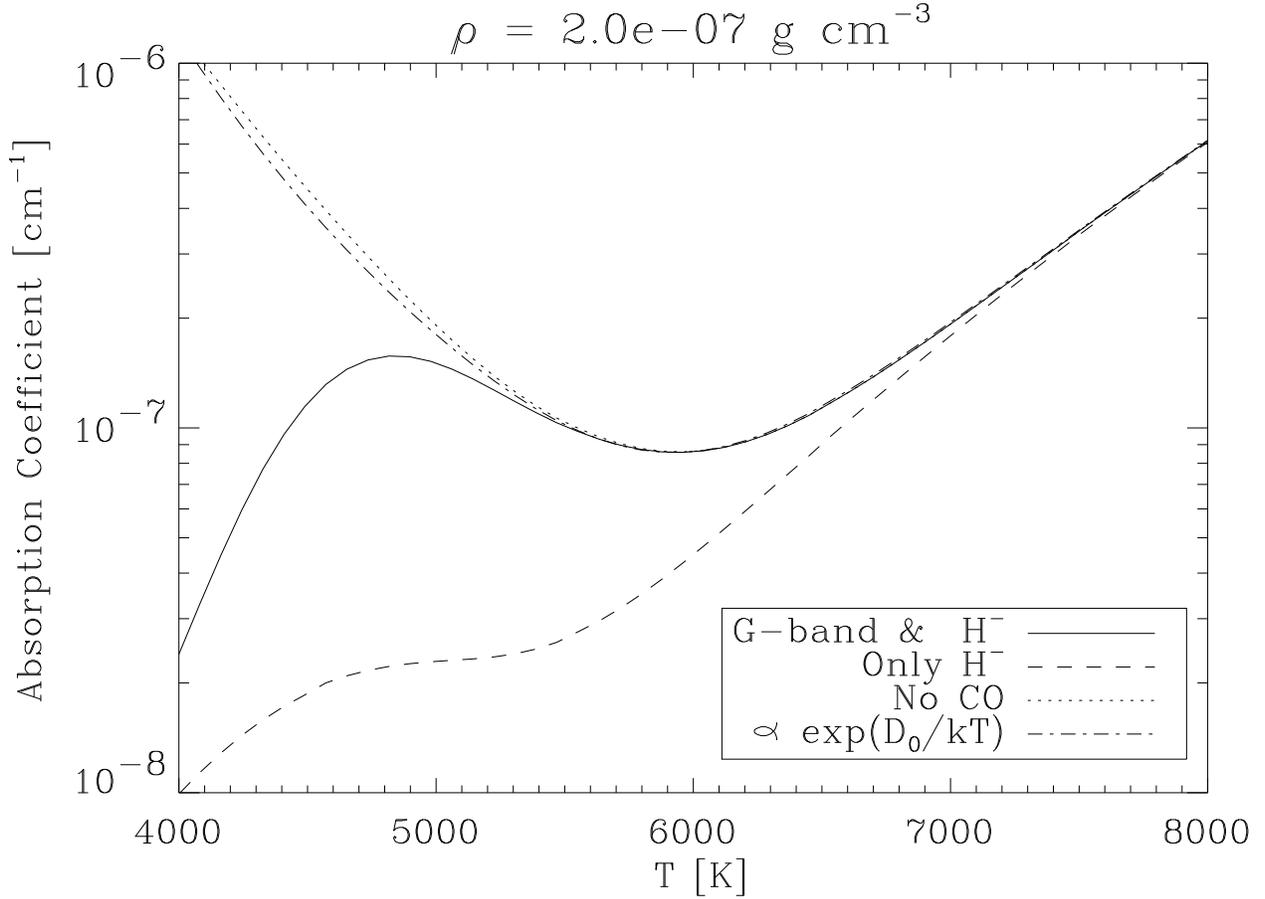}
\caption{
	Variation with temperature of the wavelength mean absorption coefficient.
	The dashed line represents the continuum absorption coefficient
	(mostly due to H$^{-}$). The solid line includes both the \gband\ 
	absorption and the continuum, and it reaches a minimum at some 5900 K.
	The existence of this extreme 
	explains the large contrast observed in the \gband\ (see text).
	The dotted line also represents the total absorption (\gband\ plus
	continuum), except that the formation of CO is turned off.
	As for the dash dot line, the 
	\gband\ opacity is approximated by the Boltzmann factor  
	of the dissociation equilibrium. The mass density $\rho$ has been
	set to a typical photospheric value (2 $\times\ 10^{-7}$
	g cm$^{-3}$).
\label{physics}}
\end{figure}

Our purpose with this work is not just explaining
the
empirical  relationship 
between magnetic concentrations and \gband\ BPs.
Actually, we find it more interesting to find  a 
reliable physical connection
between the two phenomena that can be employed 
to build on. 
Once the physics of the association
is well established, one can use modeling 
to examine the real diagnostic capabilities of the 
\gband\ BPs.  Some of the calculations  described
in this paper
begin to explore such possibility, and they already lead to
several interesting results. We condense them
here since they are scattered in  
main text.
The \gband\ signals come from deep
photospheric layers that barely differ from the
layers that produce continuum photons (\S \ref{hofs}
and Figure \ref{cfs}).  
As it was advanced  by \citeN{rut99}, 
the \gband\ opacity  seems to
be controlled by the dissociation of the CH molecule
(see \S \ref{why}). 
The agent that determines this dissociation is not
the radiation field, though. The densities in the
photosphere are so high that the
CH abundance is set by collisions with
hydrogen (reactions [\ref{destruction}] and [\ref{creation}]; see
\S \ref{photod}), 
photo-dissociation being secondary.
Since the \gband\ results from absorption of
saturated spectral lines,  it very much depends on 
the line broadening parameters characterizing the atmosphere.
Moderate
changes of micro-turbulence and damping
modify the \gband\ contrast.  For large enough turbulence,
magnetic concentrations may even show  neutral contrast in the
\gband\ (\S \ref{ltevsobs}). It is therefore
plausible that some of the real solar magnetic concentrations do not
show up in \gband\ images.
On the contrary, the continuum contrast does not
depend on line broadenings and the magnetic concentrations will tend to
be bright
in continuum.  We hypothesize the existence of low contrast continuum BPs without
a \gband\ counterpart\footnote{\citeN{san00c} points out that the
existence of a BP may be 
even better than polarization to indicate the
existence of a magnetic concentration.
According to the simulations
carried out here, this applies to the bright points at continuum
wavelengths but not necessarily in the \gband . 
} 
(\S \ref{ltevsobs}).
The LTE syntheses predict intrinsic
\gband\ signals  several times larger than those presently
observed (\S \ref{ltevsobs}). Consequently, we foresee an important augment of contrast
upon improvement of the angular resolution of the observations.
Finally, both single fluxtubes and MISMA model magnetic concentrations
produce similar \gband\ brightness. This observable does not help
discriminating between the different options of modeling unresolved
magnetic concentrations.

\modification{
We have shown that the synthetic LTE \gband\ spectra
fulfil many observed properties of the
\gband\ BPs. We have not shown that there is no 
other
way to reproduce them.
In particular, 
various alternatives mentioned in \S \ref{introduction} still
remain to be explored and
therefore cannot be discarded.
}

	While this paper was in preparation, we knew of two other
	groups carrying out LTE \gband\ syntheses in magnetic concentrations.
	They have advanced some results in recent meetings (e.g., \citeNP{rut01b};
	\citeNP{ste01}).
	As far as the overlapping with the present work is concerned,
	they find that specific model concentrations  based on single fluxtubes
	present a \gband\ contrast larger than the continuum contrast.
	The values of these contrasts are similar to those computed
	in \S \ref{ltevsobs}.

\acknowledgments
Thanks are due to H. Uitenbroek for sending us his
stellar atmosphere code, which we have employed for testing purposes.
Discussions with R. Rutten on the origin of the G-band
were extremely useful. He deserves credit for
spurring the present interest for understanding 
the \gband\ formation. 
\modification{
We  also acknoledge the detailled revision of the
manuscript carried out by an anonymous referee, from which
we have benefited.
}
This work has been partly funded by the Spanish DGES under projects
95-0028-C, PB96-0883 and ESP98-1351E. It has been carried out within 
the EC-TMR European  Solar Magnetometry Network.

\clearpage

%% No more than seven \figcaption commands are allowed per page,
%% so if you have more than seven captions, insert a \clearpage

%% There must be a \figcaption command for each legend. Key the text of the
%% legend and the optional \label in curly braces. If you wish, you may
%% include the name of the corresponding figure file in square brackets.
%% The label is for identification purposes only. It will not insert the
%% figures themselves into the document.
%% If you want to include your art in the paper, use \plotone.
%% Refer to the on-line documentation for details.
%
%


%%%%%%%%%%%%%%%%%%%%%%
%
% 	Tables

\clearpage

% New summary table
\begin{deluxetable}{ccccc}
\tablecaption{Role of 
	CH photo-dissociation in setting the chemical equilibrium.\label{table0}}
\tablewidth{0pt}
\tablehead{\colhead{T [K]}&
	\colhead{$\alpha_{photo}$ [s$^{-1}$]}&
	\colhead{$ \overline{t} $ [s]} &
	\colhead{$ n_{H} $ [cm$^{-3}$]}&
	\colhead{$r_{photo}$\tablenotemark{a}}
	}
\startdata
4400& $1.65\times 10^{2}$ & $6.1 \times 10^{-3}$ & $2.7 \times 10^{15}$ & $4.8 \times 10^{-4}$\\
4400& $1.65\times 10^{2}$ & $6.1 \times 10^{-3}$ & $1.2 \times 10^{17}$ & $1.1 \times 10^{-5}$\\
6520& $4.05\times 10^{3}$ & $2.5 \times 10^{-4}$ & $2.7 \times 10^{15}$ & $1.2 \times 10^{-2}$\\
6520& $4.05\times 10^{3}$ & $2.5 \times 10^{-4}$ & $1.2 \times 10^{17}$ & $2.6 \times 10^{-4}$
\enddata
\tablenotetext{a}{
	Probability of CH destruction by 
	photo-dissociation as defined in
		equation (\ref{ratio}).}
\end{deluxetable}

%
%% End of file 

\end{document}